\RequirePackage{fix-cm}
\documentclass[smallextended]{svjour3}       
\smartqed  
\usepackage{graphicx}
\usepackage{caption}
\usepackage{amsfonts}
\usepackage{amsmath}
\usepackage{booktabs}
\usepackage{siunitx}
\usepackage[table,dvipsnames]{xcolor}
\usepackage{hyperref}
\usepackage{cite}
\usepackage{subfigure}

\begin{document}

\title{Topic Recommendation for Software Repositories using Multi-label Classification Algorithms}
\titlerunning{Topic Recommendation for Software Repositories}
\author{Maliheh Izadi, Abbas Heydarnoori, Georgios Gousios}

\institute{M. Izadi \at
              Sharif University of Technology, Tehran, Iran. \\
              \email{maliheh.izadi@sharif.edu}\\
          \and
          A. Heydarnoori \at
              \email{heydarnoori@sharif.edu}\\
         \and
          G. Gousios \at
          The work was carried out before the author joined Facebook Inc.\\
              \email{gousiosg@fb.com}\\    
}

\date{Received: date / Accepted: date}

\maketitle

\begin{abstract}
Many platforms exploit collaborative tagging 
to provide their users with faster and more accurate results 
while searching or navigating. 
Tags can communicate different concepts such as
the main features, technologies, functionality,
and the goal of a software repository. 
Recently, GitHub has enabled users 
to annotate repositories with topic tags. 
It has also provided a set of featured topics, 
and their possible aliases, 
carefully curated with the help of the community. 
This creates the opportunity to use this initial seed of topics
to automatically annotate all remaining repositories, 
by training models that recommend high-quality topic tags to developers.

In this work, we study the application of 
multi-label classification techniques 
to predict software repositories' topics.
First, we map the large-space of user-defined topics to those featured by GitHub.
The core idea is to derive more information from projects' available documentation.
Our data contains about $152$K GitHub repositories and $228$ featured topics.
Then, we apply supervised models on repositories' textual information
such as descriptions, README files, wiki pages, and file names.

We assess the performance of our approach 
both quantitatively and qualitatively.
Our proposed model achieves 
Recall@5 and LRAP scores of $0.890$ and $0.805$, respectively.
Moreover, based on users' assessment, 
our approach is highly capable of recommending 
correct and complete set of topics.
Finally, we use our models to develop an online tool
named \texttt{Repository Catalogue}, 
that automatically predicts topics for GitHub repositories
and is publicly available\footnote{\url{https://www.repologue.com/.}}

\end{abstract}

\section{Introduction} \label{sec:intro}
Open-source software (OSS) communities provide 
a wide range of functional and technical features 
for software teams and developers 
to collaborate, share, and explore software repositories. 
Many of these repositories are similar to each other, 
i.e., they have similar objectives, 
employ common technologies 
or implement similar functionality. 
Users explore these repositories to search for 
interesting software components tailored to their needs.
However, as the community grows, 
it becomes harder to effectively organize repositories
so that users can efficiently retrieve and reuse them.

Collaborative tagging has significantly impacted 
the information retrieval field for the better, 
and it can be a promising solution to the above problem \cite{wang2018entagrec++}. 
Tags are a form of metadata used to 
annotate various entities based on their main concepts. 
They are often more useful compared to textual descriptions 
as they capture the salient aspects of an entity in a simple token. 
In fact, through encapsulating human knowledge, 
tags help bridge the gap between technical 
and social aspects of software development \cite{treude2009tagging}.
Thus, tags can be used for organizing and searching for software repositories as well.
Software tags describe categories a repository may belong to, 
its main programming language, the intended audience, 
the type of user interface, and its other key characteristics. 
Furthermore, tagging can link topic-related repositories to each other 
and provide a soft categorization of the content \cite{wang2018entagrec++}. 
Software repositories and QA platforms 
rely on users to generate and assign tags to software entities.
Moreover, several studies have exploited tags 
to build recommender systems for software QA platforms such as Stack Overflow \cite{xia2013tag,wang2014tag,wang2018entagrec++,liu2018fasttagrec}.

In $2017$, GitHub enabled its users to assign topic tags to repositories. 
We believe topic tags, 
which we will refer to as ``topics'' in this paper, 
are a useful resource for training models 
to predict high-level specifications of software repositories.
However, as of February $2020$,
only $5\%$ of public repositories in GitHub 
had at least one topic assigned to them\footnote{Information retrieved using GitHub API.}.  
We discovered over $118$K unique \textit{user-defined} topics in our data.
According to our calculations, 
the majority of tagged repositories only have 
a limited number of high-quality topics. 
Unfortunately, as users keep creating and assigning new topics 
based on their personalized terminology and style, 
the number of defined topics explodes, and their quality degrades \cite{golder2006usage}. 
This is because tagging is a distributed process, 
with no centralized coordination.
Thus, similar entities can be tagged differently \cite{xia2013tag}. 
This results in an increasing number of redundant topics 
which consequently makes it hard to retrieve similar entities 
based on differently-written synonym topics. 
For example, the same topic can be written in full or abbreviated, plural or singular formats, 
with/without special characters such as `-', 
or may contain human-language related errors, such as typos. 
Take repositories working on a deep learning model named 
\textit{Convolutional Neural Network} as an example.
We identified $16$ differently-written topics or combination of separate topics for this concept including 
\texttt{cnn}, 
\texttt{CNN},
\texttt{convolutional-neural-networks}, 
\texttt{convolutionalneuralnetwork}, 
\texttt{convolutional-deep-learning}, 
\texttt{ccn-model},
\texttt{cnn-architecture},
and \texttt{convolutional} + \texttt{neural} + \texttt{network}.
The different forms of the same concept are called \textit{alias}es. 
This high level of redundancy and customization can adversely
affect information retrieval tasks.
That is, the quality of topics (e.g., their conciseness, completeness, and consistency), 
impacts the efficacy of operations that rely on topics to perform.
Fortunately, GitHub has recently provided a set of refined topics called \textit{featured topics}.
This allows us to use this set 
as an initial seed to train supervised models to automatically tag software repositories 
and consequently, create an inventory of them.

We treat the problem of assigning existing topics to new repositories
as a multi-label classification problem.
We use the set of featured topics as labels for supervising our models. 
Each software repository can be labeled with multiple topics. 
More specifically, in the first task, 
we map the large-space of user-defined topics to their corresponding featured topics 
and then evaluate this data component.
In the second task, we use both traditional machine learning techniques 
and advanced deep neural networks, 
to train different models for automatically predicting these topics.
The input to our model consists of various types of information namely, 
a repository's name, description, README files, wiki pages, and finally its file names.
Recommender systems return ranked lists of suggestions. 
Thus, our model outputs 
a fixed number of topics with the highest predicted probabilities for a given repository.

We aim at answering the following research questions 
to address different aspects of both our data component and the classifier models:
\begin{itemize}
    \item \textbf{RQ1}: How well can we map user-defined topics to their corresponding featured topics?
    \item \textbf{RQ2}: How accurately can we recommend topics for repositories?
    \item \textbf{RQ3}: How accurate and complete are 
    the set of recommended topics from users' perspective?
    \item \textbf{RQ4}: Does the combination of input types 
    actually improve the accuracy of the models?
\end{itemize}

We first define a set of heuristic rules 
to automatically clean and transform user-defined topics through several text processing steps.
After each step, 
we manually check the results and update the rules if necessary.
Subsequent to obtaining 
the mapped dataset of user-defined and featured topics,
we perform a human evaluation 
to assess the quality and accuracy of these mappings in RQ1.
The results indicate that we are able to accurately map these topics with $98.6\%$ success rate.
In answering RQ2, 
we evaluate the performance of our models for topic recommendation based on 
various metrics including 
$R@n$, $P@n$, $F1@n$, $S@n$, and Label Ranking Average Precision ($LRAP$) scores 
of the recommended lists.
The results indicate that our approach can achieve 
high Recall, Success Rate, and LRAP scores ($0.890$, $0.971$, and $0.805$ respectively).
We also improve upon the baseline approach by $59\%$, $65\%$, $63\%$, $29\%$ and $46\%$ 
regarding $R@5$, $P@5$, $F1@5$, $S@5$ and $LRAP$ metrics, respectively.

To answer RQ3, 
we compare the recommendations of our model 
with those of the baseline approach from users' perspectives. 
Participants evaluated the recommendations based on 
two measure of \textit{correctness} and \textit{completeness}. 
Our model on average recommends $4.48$ correct topics 
out of $5$ topics for sample repositories,
while the baseline only suggests $3$ correct topics on average.
Moreover, developers indicated our model also provides 
a more complete set of recommendations compared to those of the baselines.
Finally, with RQ4, we aim at investigating the necessity of different parts of our input data.
We feed the models with different combinations of input types 
and evaluate the performance on the two best models.
The results show adding each type of information boosts the performance of the model.
Finally, our main contributions are as follows:
\begin{itemize}
    \item We perform rigorous text processing techniques 
    on user-defined topics and map $29$K of them to the GitHub's
    initial set of $355$ featured topics;
    We also assess the quality of the these mappings using human evaluation.
    \item We train several multi-label classification models 
    to automatically recommend topics for repositories.
    Then, we evaluate our proposed approach both quantitatively and qualitatively.
    The results indicate that we outperform the baseline in both cases by large margins.
    \item We make our models and datasets publicly available for use by others\footnote{\url{https://github.com/MalihehIzadi/SoftwareTagRecommender}}.
    \item Finally, we develop an online tool, \texttt{Repository Catalogue}, 
    to automatically predict topics for GitHub repositories.
    Our tool is publicly available at \url{https://www.repologue.com/.}
\end{itemize}


\section{Problem Definition}\label{sec:problem}
\begin{figure}
    \centering
    \includegraphics[width=\textwidth]{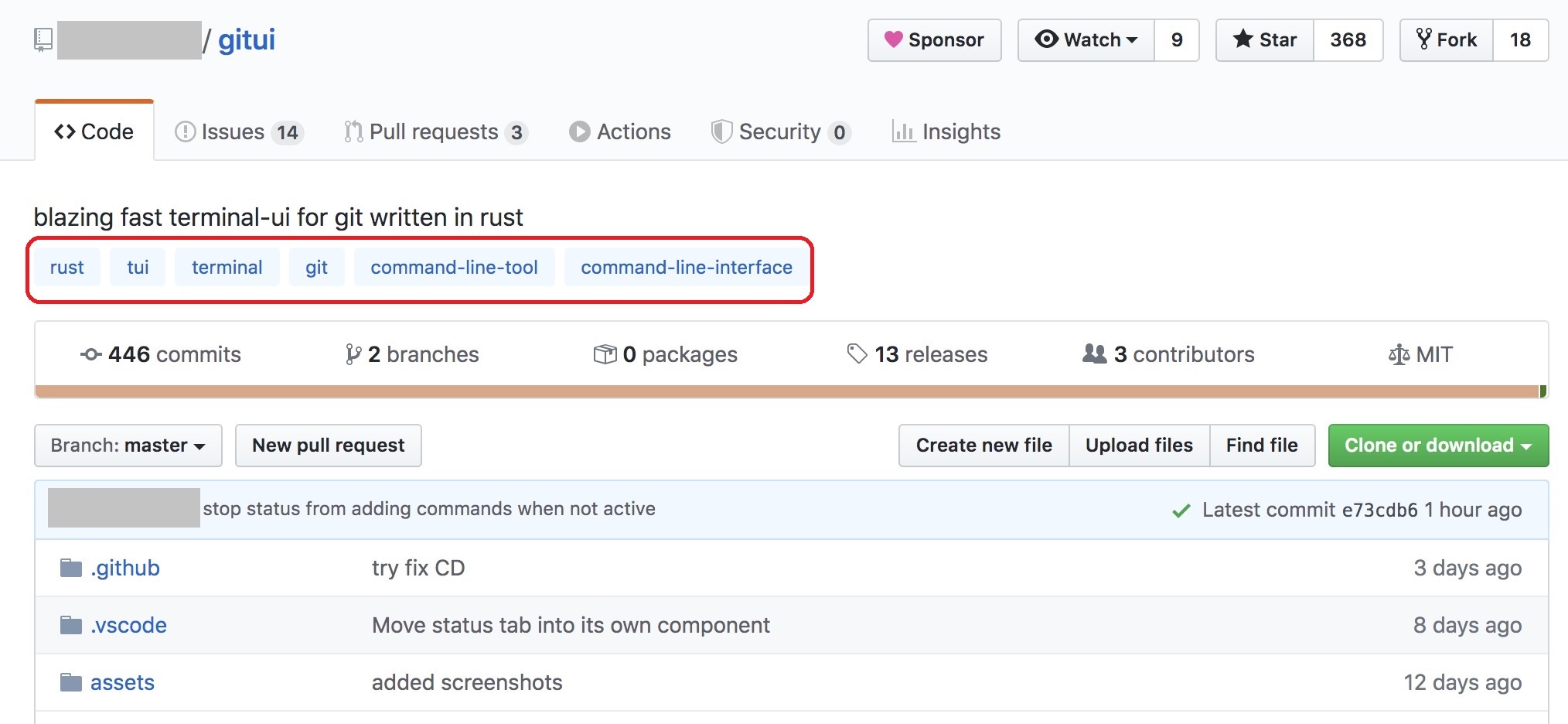}
    \caption{A sample repository and its topics}
    \label{fig:sample_repo}
\end{figure}

An OSS community such as GitHub 
hosts a set of repositories
$S = \{r_1, r_2, .., r_n\}$, 
where $r_i$ denotes a single software repository.
Each software repository may contain various types of textual information 
such as a description, README files, and wiki pages 
describing the repository's goal, and features in detail. 
It also contains an arbitrary number of files 
including its source code.
Figure \ref{fig:sample_repo} provides a sample repository from GitHub
which is tagged with six topics such as \texttt{rust} and \texttt{tui}. 
We preprocess and combine the textual information of these repositories, 
such as their name, description, README file, and wiki pages 
with the list of their file names as the input of our approach. 
Furthermore, we preprocess their set of user-defined topics,
map them to their corresponding featured topic and then
use them as the labels for our supervised machine learning techniques. 
Topics are transformed according to the initial candidate set of topics
$T = \{t_1, t_2, ...,t_m\}$, 
where $m$ is the number of featured topics. 
For each repository, $ t_i$ is either $0$ or $1$, 
and indicates whether the $i$-th topic is assigned to the target repository.
Our goal is to recommend several topics 
from the candidate set of topics $T$ to each repository $r_i$ 
through learning the relationship between 
existing repositories' textual information and their corresponding set of topics.

\section{Data Collection} \label{sec:data}
We collected the raw data of repositories 
with at least one user-defined topic using the GitHub API
which resulted in about two million repositories. 
This data contains repositories' various document files such as 
description, README files (crawled in different formats, e.g., 
\texttt{README.md}, \texttt{README}, \texttt{readme.txt}, \texttt{readme.rst}, etc.
in both upper and lower case characters), 
wiki pages, a complete list of their file names, 
and finally the project's name.
We also retrieved the set of user-defined topics for these repositories.

Initially, we remove repositories with no README and no description.
We also exclude repositories in which 
more than half of the README and description consist of non-English characters.
Then, we discard repositories that have less than ten stars \cite{kalliamvakou2016depth}.
This results in about $180$K repositories and $118$K unique user-defined topics.
After performing all the required preprocessing steps 
(Sections \mbox{\ref{sec:preprocess_topics}}, \mbox{\ref{sec:preprocess_readme}} 
and \mbox{\ref{sec:preprocess_files}}), 
we remove repositories that are left with no input data 
(either textual information or cleaned topics).
Therefore, about $152$K repositories and $228$ featured topics remains in the final data.

Considering the differences in our input sources, 
we treat textual information from these resources differently. 
We review all the preprocessing steps in more detail in their respective sections;
preprocessing topics in Section \mbox{\ref{sec:preprocess_topics}}, 
cleaning input textual information such as descriptions, READMEs, and wiki pages in Section \mbox{\ref{sec:preprocess_readme}}, 
and finally preprocessing project and file names in Section \mbox{\ref{sec:preprocess_files}}.

\section{Mapping User-defined Topics}
GitHub provides a set of community-curated topics on-line\footnote{\url{https://github.com/github/explore/tree/master/topics}.
Each of these topics may have several aliases as well. 
On February $2020$, GitHub provided 
a total number of $355$ featured topics along with $777$ aliases.
Among our $180$K repositories, 
about $136$K repositories contain at least one featured topic.
However, our dataset also contains $118$K unique user-defined topics 
and the number of aliases for these featured topics is very limited.}

The magnitude of number of user-defined topics is due to the fact that 
topics are written in free-format text. 
For instance, topics could be written as their abbreviation/acronym 
or in their full form, in plural or singular, 
with or without numbers (denoting version, date, etc.), 
and with numbers in digits or letters.
Moreover, the same topic can take different forms 
such as having ``ing" or ``ed" at its end.
Some users include stop words in their topics, some do not. 
Some have typos.
Some include words such as 
\texttt{plugin}, \texttt{app}, \texttt{application}, etc. in one topic (with or without a dash).
Note that topics written in different lexicons 
can represent the same concepts.
Furthermore, a topic that has different parts, if split, 
can represent completely different concepts compared to 
what it was originally intended to represent.
For example, \texttt{single-page-application} 
as a whole represents a website design approach. 
However, if split, part of the topic such as \texttt{single} 
may lose its meaning or worse, become misguiding.

To address the above issues, 
we preprocess user-defined topics 
and map them to their respective featured topics. 
The goal is to 
(1) exploit the large-space of user-defined topics 
by mapping them to their corresponding GitHub's featured set and 
(2) provide as many properly-labeled repositories 
as possible for the models to train and mitigate the sparsity in the dataset.
In doing so, we are able to map $29$K of user-defined topics 
to one of the $355$ featured topics of GitHub.
To assess the accuracy of our mappings, 
we design a human evaluation and evaluate the accuracy of our mappings.
In the following, we provide more details on the mapping of topics.

\subsection{Preprocessing User-defined Topics}\label{sec:preprocess_topics}
To clean and map the user-defined topics, 
we extract existing featured topic from the list of user-defined topics (if any) as the first step.
Then, we use a set of heuristics, 
and perform the following text processing steps on user-defined topics.
After each step, two of the authors manually inspect the results, 
and update the rules if necessary.
\begin{itemize}
    \item Remove versionings, 
    e.g., \texttt{v3} is removed from \texttt{react-router-v3},
    \item Remove digits at the end of a topic, 
    e.g., \texttt{php7} is changed to \texttt{php}
    (note that we cannot simply remove any digits since topics, 
    such as \texttt{3d}, and \texttt{d2v} will lose their meaning),
    \item Extract the most frequent topics 
    such as \texttt{api}, \texttt{tool}, or \texttt{package} 
    from the rest of user-defined topics. For example, \texttt{twitch-api} is converted to two separate topics of \texttt{twitch} and \texttt{api},
    \item Convert plural forms to singular,
    e.g., \texttt{components} is converted to \texttt{component}.
    Note that one cannot simply remove `s' 
    from the end of a topic because topics such as
    \texttt{js}, \texttt{css}, \texttt{kubernetes}, \texttt{iOS} 
    will become meaningless),
    \item Replace abbreviations,
    e.g., \texttt{os} is expanded to \texttt{operating-system} 
    and \texttt{d2v} is converted to \texttt{doc2vec}.
    \item Remove stop words such as \texttt{of}, and \texttt{in},
    \item Lemmatize topics to preserve the correct word form: 
    for instance \texttt{reproducer} is converted to \texttt{reproduce},
    \item Aggregate topics. 
    For this step, two of the authors manually identified 
    a set of topics that when aggregated can represent a larger concept. 
    For example, for repositories tagged with both 
    \texttt{neural} and \texttt{network} topics, 
    we combine these two topics 
    and merge them into one main topic of \texttt{neural-network}. 
    Other examples include bigrams such as
    \texttt{machine} and \texttt{learning}, \texttt{package} and \texttt{manager}, 
    or trigrams such as \texttt{windows}, \texttt{presentation}, and \texttt{foundation}.
    The complete list is available in our repository.
\end{itemize}

\begin{figure}
    \centering
    \includegraphics[width=1\textwidth]{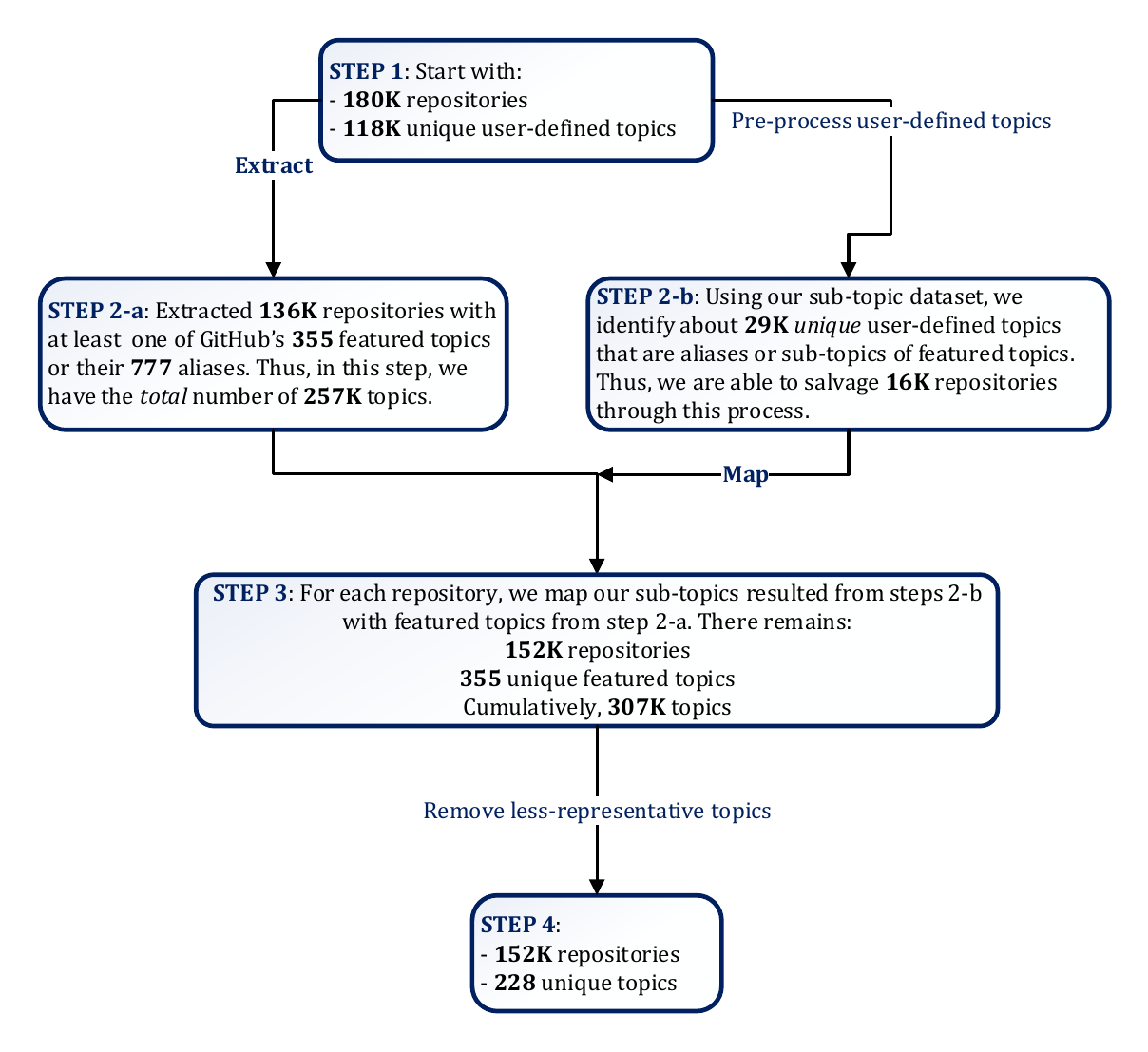}
    \caption{Mapping user-defined topics to featured topics}
    \label{fig:tags_preprocess}
\end{figure}

After cleaning and transforming user-defined topics according to the above, 
we obtain a set of mapped sub-topics to their corresponding featured topics.
Next, we augment the set of a repository's featured topics 
(output of the first step)
with our set of the mapped featured topics 
(recovered from the rest of above steps). 
Figure \ref{fig:tags_preprocess} depicts 
the process of mapping the sub-topics with their featured versions.
We discovered about $29$K unique sub-topics 
that can be mapped to their corresponding featured topics.
Furthermore, we recover $16$K more repositories (from our $180$K repositories)
and increase the total number of featured topics used in the dataset by $20\%$.
In this stage, data contains about $152$K repositories with $355$ unique featured topics 
and total of $307$K tagged featured topics.
To have sufficient number of sample repositories both in the training and testing sets,
we remove the less-represented feature topics (used in less than $100$ repositories in the dataset). 
There remains a set of $228$ featured topics.

It is worth mentioning that while GitHub provides 
on average two aliases for each featured topic,
we were able to identify on average $94$ sub-topics for each featured topic.
Moreover, while Github does not provide any alias for $95$ featured topics,
we were able to recover at least one sub-topic for half of them ($48$ out of $95$).
Table \ref{tab:subtopics_stat} summarizes the statistics information 
about GitHub's aliases and our sub-topics per repository.
Table \ref{tab:sample_repo_subtopics} presents a sample of 
GitHub repositories, their user-defined topics, 
the directly extracted featured topics, 
and the additional mapped featured topics using our approach.
In section \ref{sec:assess_data}, 
we perform a human evaluation on a 
statistically representative sample of this $29$K sub-topics dataset 
and assess the accuracy of mapped pairs of (sub-topic, featured topic,).
\begin{table*}[ht]
\caption{Statistics summary for aliases and sub-topics}
\centering
\begin{tabular}{cccccc}\toprule
    && \multicolumn{4}{c}{Per featured topic}\\
    \cmidrule{3-6}
    Source         & Unique Number & Min & Max & Mean & Median\\\midrule
    Aliases by GitHub        & $777$   & $0$ & $102$ & $2$ & $1$\\
    Our sub-topics     & $29$K & $0$ & $1860$  & $94$  & $26$ \\\bottomrule
\label{tab:subtopics_stat}
\end{tabular}
\end{table*}

\begin{table*}[ht]
\caption{Mapping user-defined topics to proper featured topics (samples)}
\centering
\begin{tabular}{p{25mm}p{30mm}p{15mm}p{30mm}}\toprule
    \textbf{Repository} 
    & \textbf{user-defined \newline topics}
    & \textbf{Extracted \newline featured topics} 
    & \textbf{Extra mapped \newline featured topics}\\
    \midrule\midrule
    \href{https://github.com/kubernetes-sigs/gcp-compute-persistent-disk-csi-driver}{kubernetes-sigs/gcp-compute-persistent-disk-csi-driver} 
    & k8s-sig-gcp, gcp 
    & -
    & google-cloud, \newline kubernetes\\
    \midrule
    \href{https://github.com/microsoft/vscode-java-debug}{microsoft/vscode-java-debug} 
    & java, java-debugger, \newline vscode-java 
    & java
    & visual-studio-code\\
    \midrule
    \href{https://github.com/fandaL/beso}{fandaL/beso}
    & topology-optimization, calculix-fem-solver, \newline finite-element-analysis
    & -
    & finite-element-method\\
    \midrule
    \href{https://github.com/mdwhatcott/pyspecs}{mdwhatcott/pyspecs} 
    & testing-tools, \newline tdd-utilities, bdd-framework, python2
    & -
    & python, testing\\
    \bottomrule
\label{tab:sample_repo_subtopics}
\end{tabular}
\end{table*}

\begin{figure}[!ht]
    \centering
    \caption{Statistical information about the dataset}
    \subfigure[Topic number per repository histogram]
    {\includegraphics[width=0.48\linewidth]{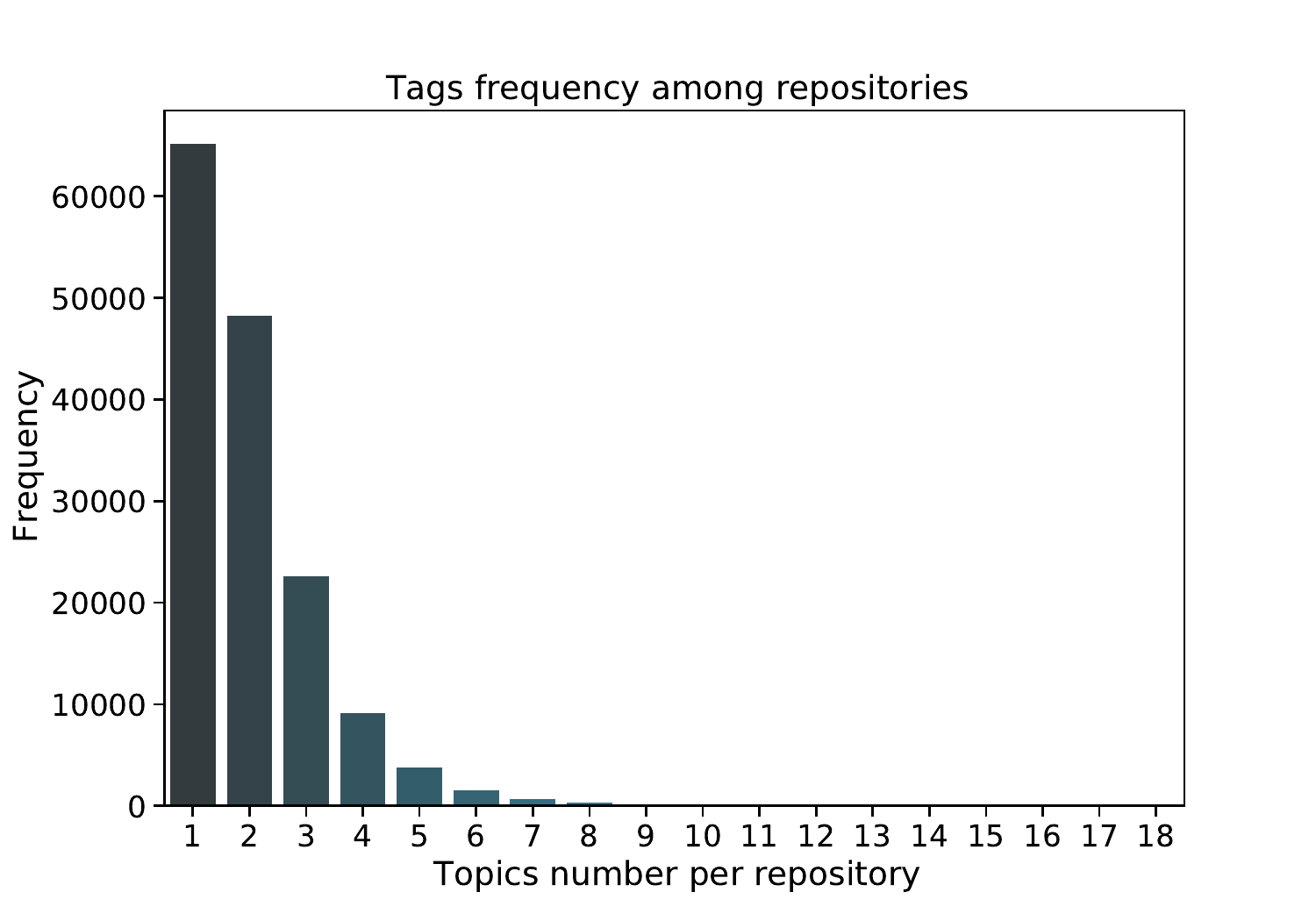}}
        \label{fig:topics_num_dist}
    \subfigure[Coverage rate of topics]
    {\includegraphics[width=0.48\linewidth]{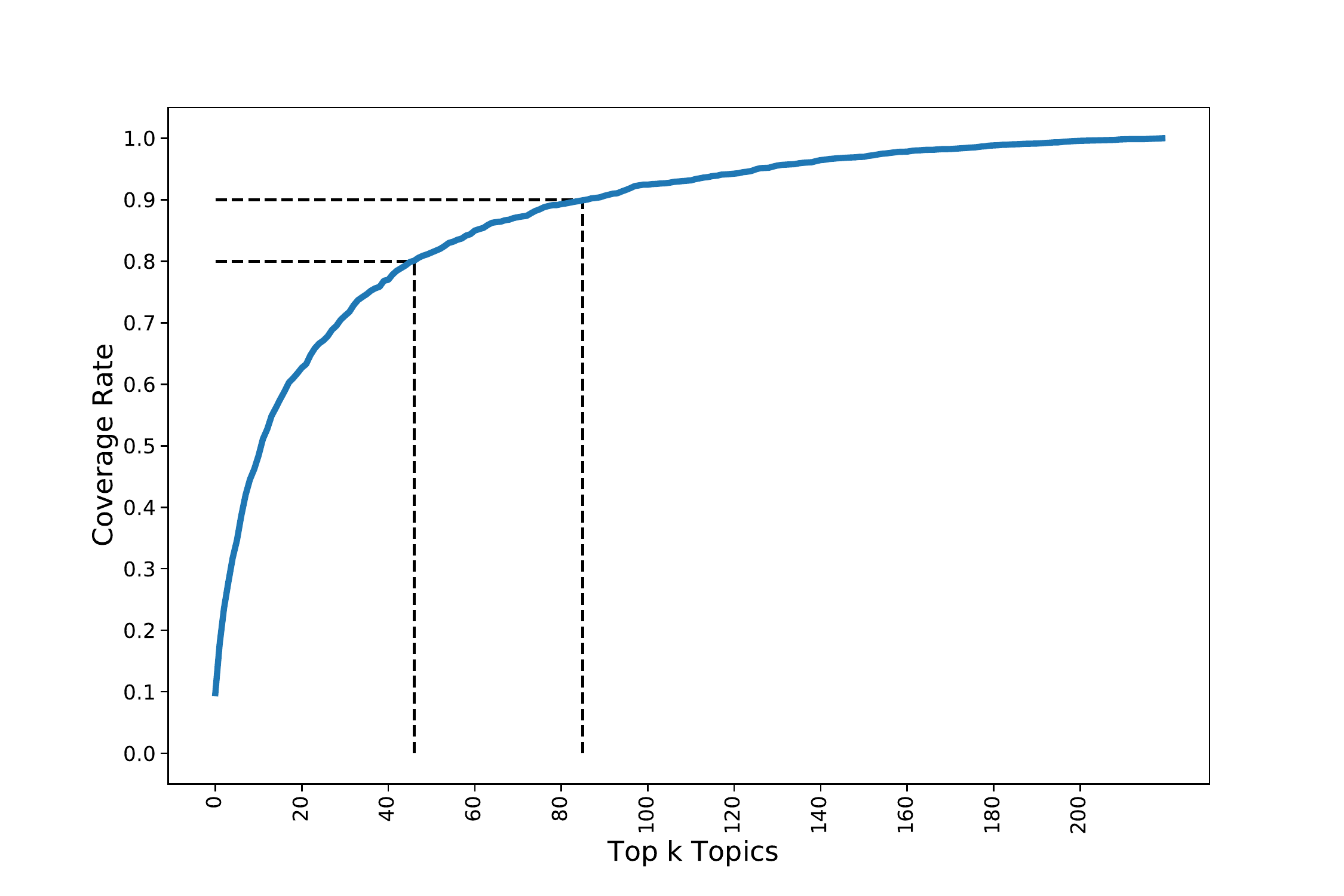}}
        \label{fig:freq_100}
    \subfigure[Most frequent 45 topics]
    {\includegraphics[width=0.7\linewidth]{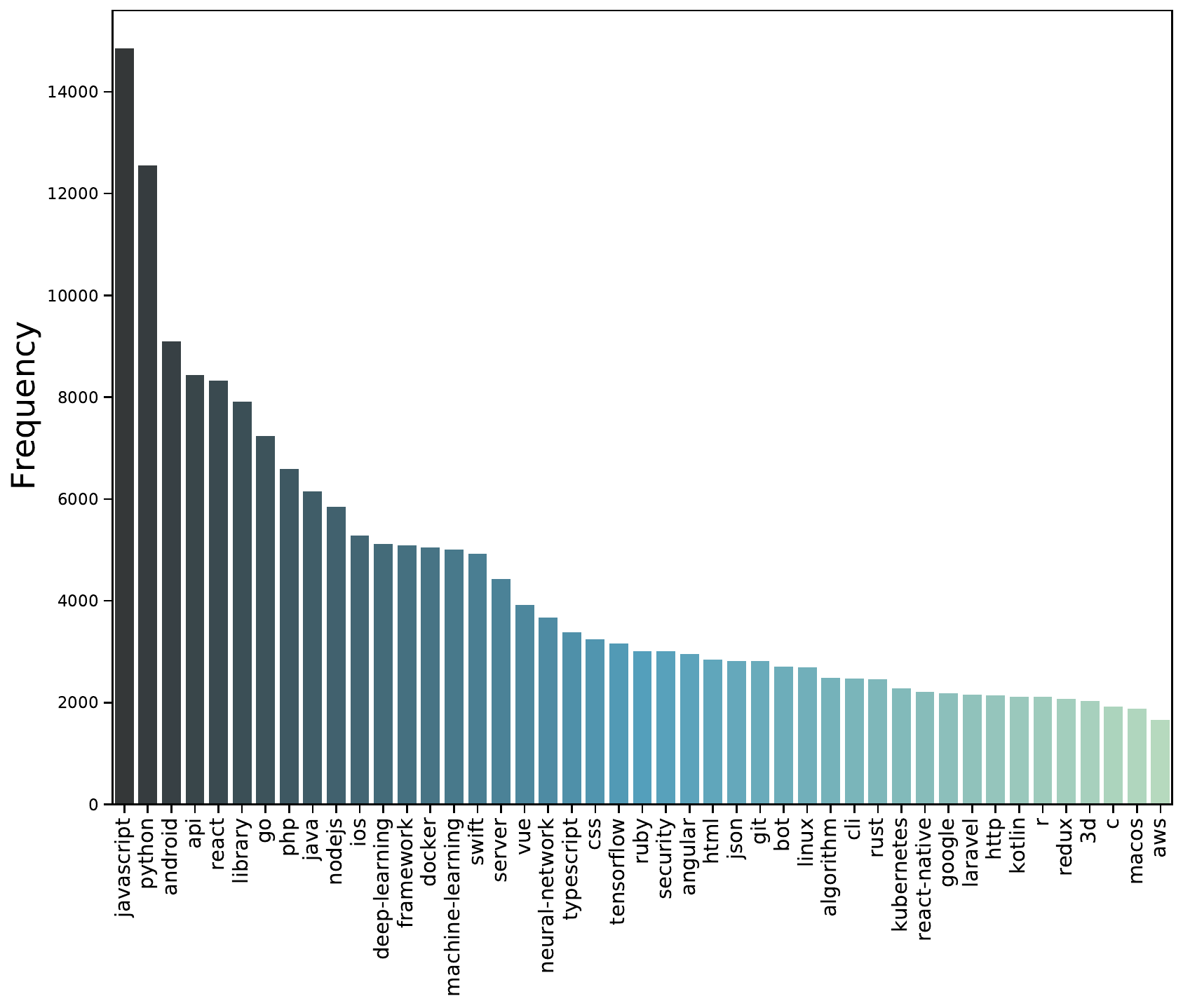}}
        \label{fig:topics_histogram}
    \label{fig:datat-stat}
\end{figure}

In the final dataset, 
almost all repositories have less than six featured topics, 
with a few outliers having up to $18$ featured topics (Figure \mbox{\ref{fig:topics_num_dist}}).
Distribution of topics among repositories has a long tail, 
i.e., a large number of them are used only in a small percent of all repositories. 
The most frequent topics are
\texttt{javascript} ($14.3$K repositories), \texttt{python} ($12.5$K),
 \texttt{android} ($8.7$K), \texttt{api}, \texttt{react}, \texttt{library}, 
 \texttt{go}, \texttt{php}, \texttt{java}, \texttt{nodejs}, \texttt{ios}, 
 and \texttt{deep-learning}. 
 The least frequent topics in our dataset are 
 \texttt{purescript}, \texttt{racket}, and \texttt{svelte}. 
Each of them were used for at least $100$ repositories.
To provide a better picture on the distribution of featured topics over the data, 
we compute topic's \textit{coverage rate}.
Equation \ref{eq:coverage_ratio} divides the sum of frequencies of 
top $k$ topics ($k$ most frequent topics) 
by sum of frequencies of all topics in the processed dataset. 
$N$ denotes the total number of topics and $frequency_i$ is the frequency of the $i$-th topic.

\begin{equation}
\label{eq:coverage_ratio}
Coverage_k = \dfrac{\sum\limits_{i=1}^{k}frequency_i}{\sum\limits_{i=1}^{N}frequency_i}
\end{equation}

As displayed in Figure \ref{fig:freq_100}, 
in our dataset, top $20\%$ number of topics cover 
more than $80\%$ of the topics' cumulative frequencies over all repositories. 
In other words, cumulative frequencies of top $45$ topics cover $80\%$  of cumulative frequencies of all topics. 
The distribution of top $45$ topics in the final dataset is shown in Figure \ref{fig:topics_histogram}. 

\subsection{Human Evaluation of the Mapping} 
To answer \textbf{RQ1}, we assessed the quality of the sub-topic dataset 
with the help of software engineering experts.
As mentioned in Section \ref{sec:preprocess_topics}, 
through cleaning $118$K user-defined topics,
we built a dataset of about $29$K unique sub-topics 
which can be mapped to the set of GitHub's $355$ featured topics.

Fourteen software engineers participated in our evaluation, 
five females and nine males.
All our participants either have 
an MSc or a PhD in Software Engineering or Computer Science.
Moreover, they have a minimum of $5.0$, 
and an average of $9.4$ years of experience 
in software engineering and programming.

As the number of sub-topics is too large for the set of topics to be manually
examined in its entirety, we randomly selected a statistically
representative sample of $7215$ sub-topics from the dataset 
and generated their corresponding pairs as \textit{(sub-topic, featured topic)}.
This sample size should allow us 
to generalize the conclusion about the success rate of the mappings
to all our pairs with a confidence level of $95\%$ and confidence interval of $1\%$.
We tried to retrieve at least $25$ sub-topics corresponding to each featured topic.
However, $47$ featured topics, 
had less number of sub-topics.

\begin{figure}
    \centering
    \includegraphics[width=0.65\textwidth]{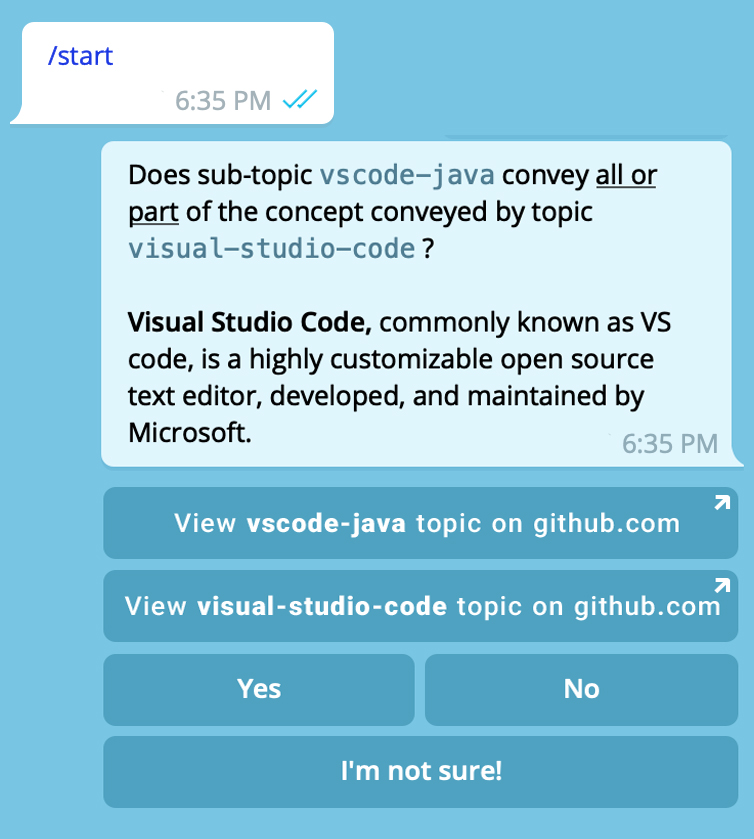}
    \caption{A screenshot of our Telegram bot}
    \label{fig:bot}
\end{figure}

We developed a Telegram bot 
and provided participants with a simple question:
``Considering the pair (featured topic $ft$, sub-topic $st$), 
Does the sub-topic $st$ convey \textit{all or part} of the concept
conveyed by the featured topic $ft$?"
to which the participants could answer 
\textit{`Yes'}, \textit{`No'}, \textit{`I am not sure!'}. 
To better provide context for the participants, 
we also included the definition of the featured topics 
and some sample repositories tagged with the sub-topic.
This would help them get a good understanding of 
definition and usage of that particular topic among GitHub's repositories.
We asked our participants 
to take their time and carefully consider each pair 
and answer with options Yes/No.
In case that they could not decide, 
they were instructed to use the `I am not sure!' button.
These cases were later analyzed 
and were labeled either as `Yes' or `No' in the final round.
For this experiment, 
we collected a minimum of two answers per pair of (featured topic, sub-topic).
We consider pairs with at least one 'No' label as failure 
and pairs with unanimous `Yes' labels as success.
Figure \ref{fig:bot} shows a screenshot of this Telegram bot.

\subsection{RQ1: Evaluating Mappings}\label{sec:assess_data}
According to the results of the human evaluation, our success rate is $98.6\%$, 
i.e., the participants confirmed that for $98.6\%$ of pairs of the sample set, 
the sub-topic was correctly mapped to its corresponding featured topic.
Only $101$ pairs were identified as failed matches.
Two of the authors discussed all the cases for which 
at least one participant had stated they believed 
the sub-topic and the featured topic should not be mapped.
After a careful round of analysis, 
incorrectly mapped topics were identified as related to 
a limited number of featured topics, namely
\texttt{unity}, \texttt{less}, \texttt{3d}, \texttt{aurelia},
\texttt{composer}, \texttt{quality}, \texttt{c}, \texttt{electron},
\texttt{V}, \texttt{code-review}, and \texttt{fish}.
For instance, we had wrongfully mapped 
\texttt{data-quality-monitoring} to \texttt{code-quality}, 
\texttt{lesscode} to \texttt{less} or 
\texttt{nycopportunity} to \texttt{unity}.
Moreover, there were also some cases where a common abbreviation 
such as \texttt{SLM} was used for two different concepts.
After performing this evaluation, we updated our sub-topic dataset accordingly.
In other words, we removed all the instances of wrong matches 
from the dataset\footnote{The updated dataset is available in our repository for public use.}.

To answer RQ1, we conclude that our approach successfully 
maps sub-topics to their corresponding featured topic.
Our participants confirmed that these sub-topics indeed 
convey a part or all of the concept conveyed by
corresponding featured topic in almost all instances of the sample set.
In the next section, we train our recommender models and evaluate the results.

\section{Topic Recommendation}\label{sec:design}
In this section, 
we first review the data preparation steps to clean the input information of the models.
Then, we present a brief background on the machine learning models,
and the high-level architecture of our approach.
Next, we discuss the main components of the approach in more detail.

\subsection{Data preparation}
Here, we preprocess our two types of input information for the models 
including repositories' description text, READMEs, wiki pages, and the file names.

\subsubsection{Preprocessing Descriptions, READMEs, and Wiki Pages}\label{sec:preprocess_readme}
We perform the following preprocessing steps on these types of data.
\begin{itemize}
    \item Remove punctuation, digits, non-English 
    and non-ASCII characters,
    \item Replace popular SE- and CS-related abbreviations and acronyms 
    such as \texttt{lib}, \texttt{app}, \texttt{config}, 
    \texttt{DB}, \texttt{doc}, and \texttt{env} 
    with their formal form in the dataset\footnote{The complete list of these tokens is available in our repository.}, 
    \item Remove abstract concepts such as 
    emails, URLs, usernames, markdown symbols, code snippets, 
    dates, and times to normalize the text using regular expressions,
    \item Split tokens based on several naming conventions 
    including SnakeCase, camelCase, and underscores 
    using an identifier splitting tool called \textit{Spiral}\footnote{\url{https://github.com/casics/spiral.}}, 
    \item Convert tokens to lower case, 
    \item Omit stop words, 
    then tokenize and lemmatize documents 
    to retain their correct word formats,
    We do not perform stemming 
    since some of our methods (e.g., DistilBERT) 
    have their own preprocessing techniques, 
    \item Remove tokens with a frequency of less than 50 
    to limit the vocabulary size for traditional classifiers. 
    Less-frequent words are typically special names or typos. 
    According to our experiments, using these tokens 
    has little to no impact on the accuracy.
\end{itemize}

\subsubsection{Preprocessing Project's and Source File Name}\label{sec:preprocess_files}
The reason for incorporating this type of information in our approach 
is that names are usually a good indicator of 
the main functionality of an entity. 
Therefore, we crawled a list of all the file names 
available inside each repository. 
As this information cannot be obtained using the GitHub API, 
we cloned every project and then parsed all their directories.
Before cleaning file names, our dataset had an average of 488 
and a median of $50$ files per repository.
We perform the following steps on the names:
\begin{itemize}
    \item Split the project name into the owner and the repository name.
    \item Drop special (e.g., `-' and `.') or non-English characters from all names, 
    \item Split names according to the naming conventions, 
    including \textit{SnakeCase}, \textit{camelCase}, and \textit{underscores} (using Spiral).
    \item Identify a list of most frequent and informative tokens such as \texttt{lib} and \texttt{api} from the list of all names 
    and split tokens based on them. For instance, \texttt{svmlib} is split to two tokens of \texttt{svm} and \texttt{lib},
    \item Omit stop words, and apply tokenization and lemmatization on the names,
    \item For the source file names, 
    remove the most frequent but not useful name tokens 
    that are common in various types of repositories 
    regardless of their topic and functionality.
    These include names such as 
    \texttt{license}, \texttt{readme}, \texttt{body}, \texttt{run}, \texttt{new}, \texttt{gitignore}, 
    and frequent file formats such as \texttt{txt}\footnote{The complete list of these tokens is available in our repository.}. 
    These tokens are frequently used 
    but do not convey much information about the topic.
    For instance, if a token such as 
    \texttt{manager} or \texttt{style}
    is repeatedly used in description or README of a repository, 
    it implies that the repository's functionality is related to these token. 
    However, an arbitrary repository can contain several files named \texttt{style} or \texttt{manager}, 
    while the repository's main functionality varies from these topics.
    Since we concatenate all the processed tokens 
    from each repository into a single document 
    and feed this document as the input to our models, 
    we removed these domain-specific tokens 
    from the list of file names to avoid any misinterpretation by the models\footnotemark.
    \item Remove tokens with a frequency of less than $20$. 
    This omits uninformative personal tokens such as user names.
\end{itemize}

\subsubsection{Statistics of Input Information}
\begin{figure}
    \centering
    \includegraphics[width=0.9\textwidth]{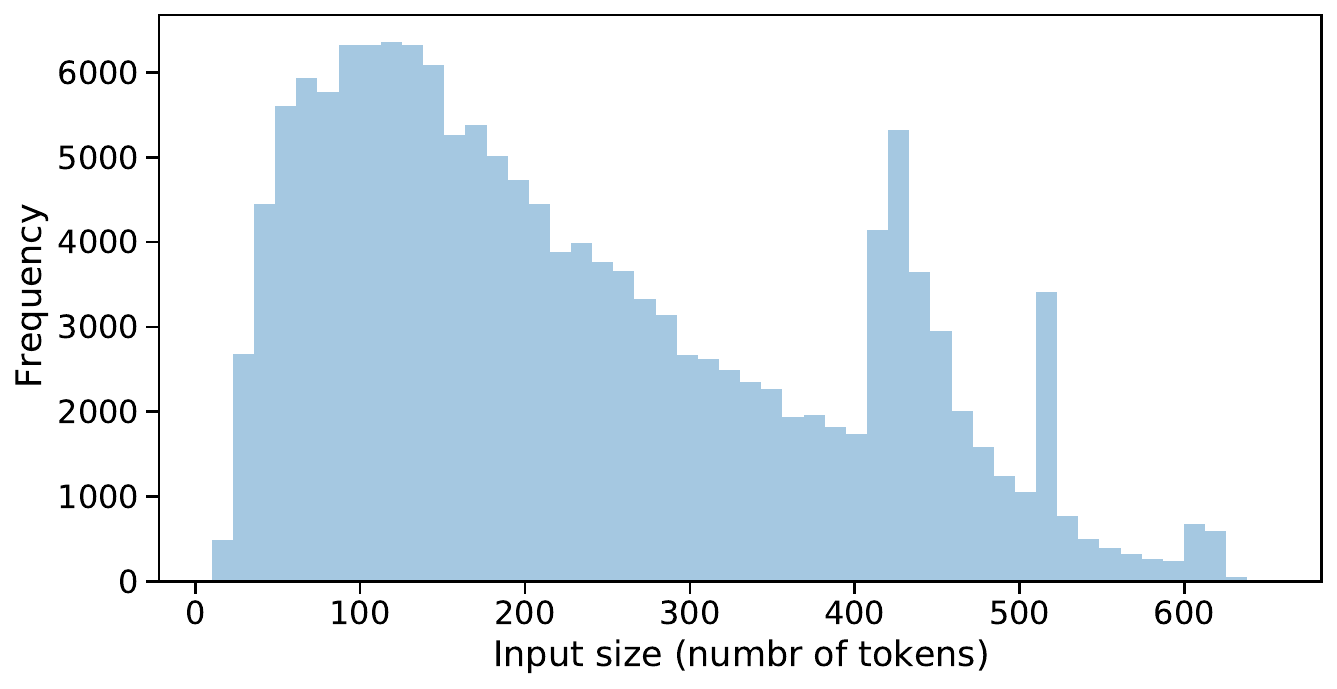}
    \caption{The histogram of input documents size based on number of tokens}
    \label{fig:input_histo}
\end{figure}

Based on the distribution of input data types, 
we truncate a fixed number of tokens 
and concatenate them to make 
a list of single input documents.
To be exact, we extract a maximum of 
$10$, $50$, $400$, $100$, and $100$ tokens 
from project names, descriptions, READMEs, wiki pages, and file names, respectively.
Some of the models can accept a limited number of input tokens, 
hence truncating the input helps us have a fair comparison.
By common assumption, the main idea is usually expressed in the opening sentences, 
thus we truncate based on the order of the tokens available in the text of descriptions, READMEs, and wiki pages.
For the file names, we start from files in the root directory and then go one level deeper in each step.
In our dataset, most of the data for each repository comes from its README files.
Figure \ref{fig:input_histo} presents a 
histogram of prevalence of the number of input tokens 
among the repositories in our dataset.
Table \ref{tab:input_stat} summarizes some statistics about our input data.
The average number of input tokens per repository is $235$.
After employing all the preprocessing steps described in previous sections, 
we concatenate all the data of each repository into a single document file 
and generate the representations for feeding to classifiers.
\begin{table*}[ht]
\caption{Input size: statistics information (152K repositories)}
\centering
\begin{tabular}{ccccc}\toprule
    & \multicolumn{3}{c}{Token number}\\
    \cmidrule{2-5}
    Source         & Min & Max & Mean & Median\\\midrule
    Project name   & 1 & 10 & 3 & 2\\
    Description    & 1 & 50 & 7 & 6\\
    README         & 1 & 400 & 175 & 140\\
    Wiki           & 1 & 100 & 10 & 1\\
    File names     & 1 & 100 & 36 & 22\\
    All            & 10 & 651 & 235 & 200\\\bottomrule
\label{tab:input_stat}
\end{tabular}
\end{table*}

\subsection{Background}
\label{sec:background}
In this section, 
we provide preliminary information on the methods we have
used in our proposed approach, 
covering both traditional classifiers and deep models.

\textbf{Naive Bayes}: 
Multinomial Naive Bayes (MNB) is a variant of Naive Bayes 
frequently used in text classification.
MNB is a probabilistic classifier used for multi-nomially distributed data. 
On the other hand, the second Naive Bayes variation, Gaussian NB (GNB),
is used when the continuous values associated with each class 
are distributed according to Gaussian distribution.

\textbf{Logistic Regression}: 
This classifier uses a logistic function to model 
the probabilities describing the possible outcomes of a single trial.

\textbf{FastText}
Developed by Facebook, 
FastText is a library for 
learning word representations and sentence classification
especially in the case of rare words 
by exploiting character level information \cite{joulin2017bag}.
We have used FastText to train a supervised text classifier.

\textbf{DistilBERT}: Transformers are the state-of-the-art models 
which exploit the attention mechanism
and disregard the recurrent component of Recurrent Neural Networks (RNN) \cite{vaswani2017attention}. 
Transformers are showed to generate higher quality results for several NLP tasks, 
they are more parallelizable, 
and require significantly less time to train compared to RNNs.
Using the transformer concept, 
Bidirectional Encoder Representations from Transformers (BERT) 
was proposed to pre-train deep bidirectional representations 
from unlabeled text by jointly conditioning on both left and right context \cite{devlin2018bert}. 
BERT employs a two tasks of Masked Language Modeling 
and Next Sentence Prediction on a large corpus 
constructed from the Toronto Book Corpus and Wikipedia.
DistilBERT developed by HuggingFace \mbox{\cite{sanh2019distilbert}}, 
was proposed to pre-train 
a smaller general-purpose language model compared to BERT. 
DistilBERT combines language modeling, distillation 
and cosine-distance losses
to leverage the inductive biases learned by pre-trained larger models.
The authors have shown DistilBERT can be fine-tuned with good performances 
on a variety of tasks. 
They claim compared to BERT, DistilBERT decreases the model size by $40\%$, 
while retaining $97\%$ of its language understanding capabilities 
and being $60\%$ faster.

\subsection{Approach Overview}
\begin{figure}
    \centering
    \includegraphics[width=1\textwidth]{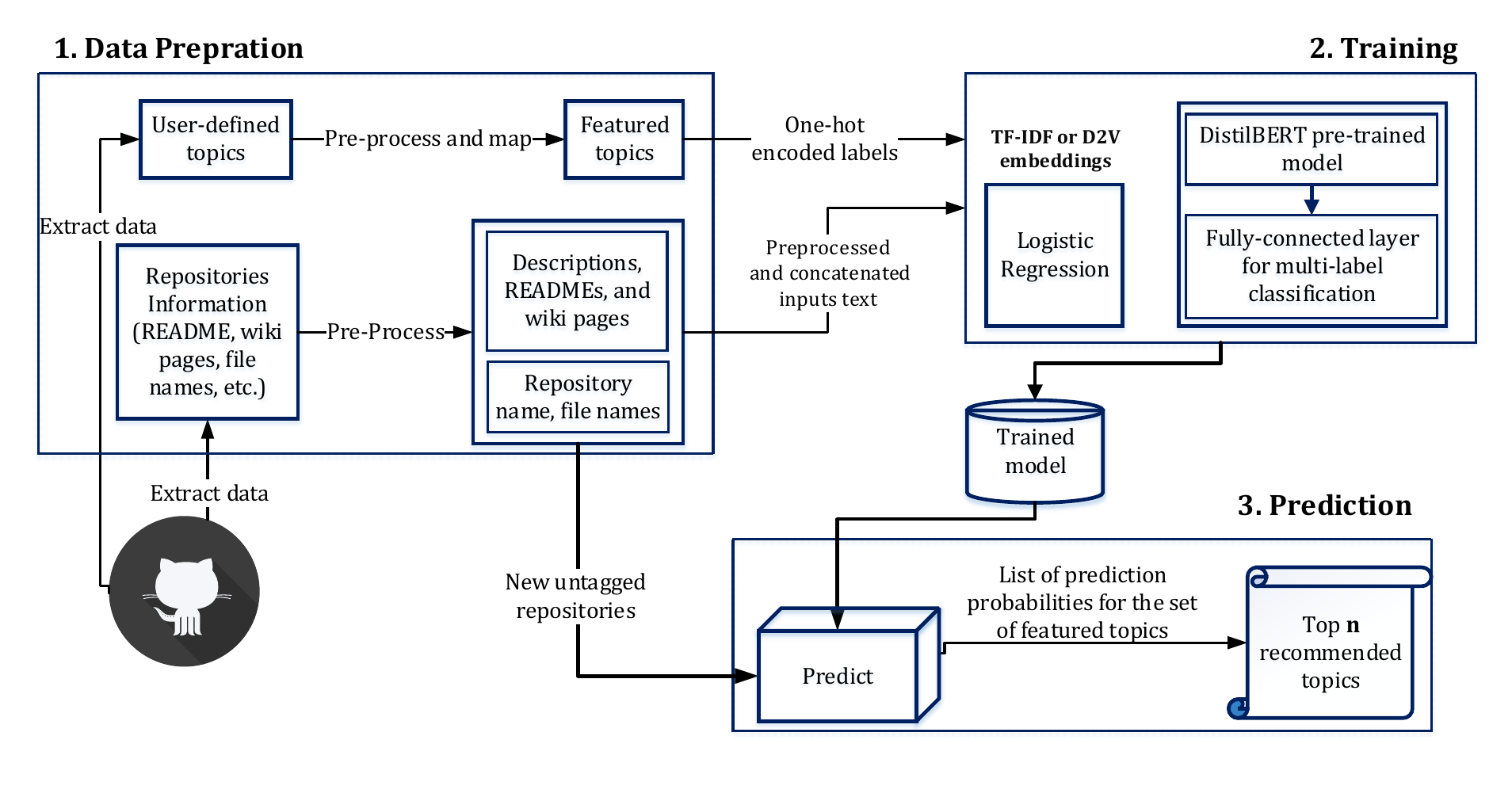}
    \caption{Overall workflow of the proposed approach}
    \label{fig:approach}
\end{figure}

Figure \ref{fig:approach} presents 
the overall workflow of our proposed approach 
consisting of three main phases; 
(1) data preparation, (2) training, and (3) prediction.

The first phase is composed of two parts;
preparing the set of featured topics  
and preparing the textual data of repositories 
as labels and inputs of the multi-label classifiers.
For each repository, we extract its available
user-defined topics, name, description, README files, wiki pages, 
and finally a list of source file names (including their extensions).
user-defined topics assigned to the repositories 
go through several text-processing steps 
and then, are compared to the set of featured topics.
After applying the preprocessing steps, 
if the cleaned version of a user-defined topic is found 
in the list of featured topics, it will be included, 
otherwise it will be discarded.
Our classifier treats the list of topics for each repository as its labels. 
We transform these featured topics' lists per repository to multi-hot-encoded vectors 
and use them in the multi-label classifiers.
We also process and concatenate 
textual data from the repositories along with their source file names 
to form our corpus.
We feed the concatenated list of a repository's textual information
(description, README, wiki, project name, and file names)
to the transformer-based and FastText classifier as is.
On the other hand, for traditional classifiers, 
we either use TF-IDF or Doc2vec embeddings
to represent the input textual information of repositories.

Next, in the training phase, 
the resulting representations are fed to the classifiers 
to capture the semantic regularities in the corpus.
The classifiers detect the relationship between 
the repositories' textual information 
and the topics assigned to the repositories 
and learn to predict the probability of each featured topic 
being assigned to the repositories.

Finally, in the prediction phase, 
the trained models predict topics for the repositories in the test dataset. 
In fact, our model output a vector containing 
probabilities of assigning each topic to a sample repository. 
We sort the output probability vector 
and then retrieve the corresponding topics for the top candidates (highest probabilities)
based on the recommendation list's size.


\subsection{Multi-label Classification}\label{sec:methods}
The classifiers we have reviewed in Section \ref{sec:background} 
are some of the most efficient and widely used 
supervised machine learning models for text classification. 
We train the following set of traditional classifiers 
with the preprocessed data acquired from the previous phase: 
MNB, GNB, and LR. 
The input data in text classification 
for these classifiers is typically represented 
as TF-IDF vectors, or Doc2Vec vectors. 
Usually, MNB variation is applied to classification problems 
where the multiple occurrences of words are important. 
We use MNB with TF-IDF vectors and GNB with Doc2Vec vectors.
We also use LR with both TF-IDF and Doc2Vec vectors.
To be comprehensive, we employ a FastText classifier as well,
which can accept multi-label input data.
As for the deep learning approaches, 
we fine-tune a DistilBERT pre-trained model to predict the topics.
We discuss our approach in more detail in the following sections.

\subsubsection{Multi-hot Encoding}
Multi-label classification is a 
classification problem where multiple target labels 
can be assigned to each observation 
instead of only one label in the case of standard classification. 
That is, each repository can have an arbitrary number of assigned topics. 
Since we have multiple topics for repositories, 
we treat our problem as a multi-label classification problem 
and encode the labels corresponding to 
each repository in a multi-hot encoded vector.
That is for each repository 
we have a vector of size $228$, 
with each element corresponding to one of our featured topics.
The value of these elements are either $0$ or $1$, 
depending on whether that repository has been assigned the target topic.

\subsubsection{Problem Transformation}
Problem transformation is an approach 
for transforming multi-label classification 
into binary or multi-class classification problems. 
OneVsRest (OVR) strategy is a form of problem transformation 
for fitting exactly one classifier per class. 
For each classifier, 
the class is fitted against all the other classes. 
Since each class is represented by only one classifier, 
OVR is an efficient and interpretable technique 
and is the most commonly used strategy 
when using traditional machine learning classifiers 
for a multi-label classification task.
The classifiers take an indicator matrix as an input, 
in which cell $[i, j]$ indicates 
that repository $i$ is assigned the topic $j$.
Using this approach of problem transformation, 
We converted our multi-label problem 
to several simple binary classification problems, 
one for each topics.

\subsubsection{Fine-tuning Transformers}
\begin{figure}
    \centering
    \includegraphics[width=0.9\textwidth]{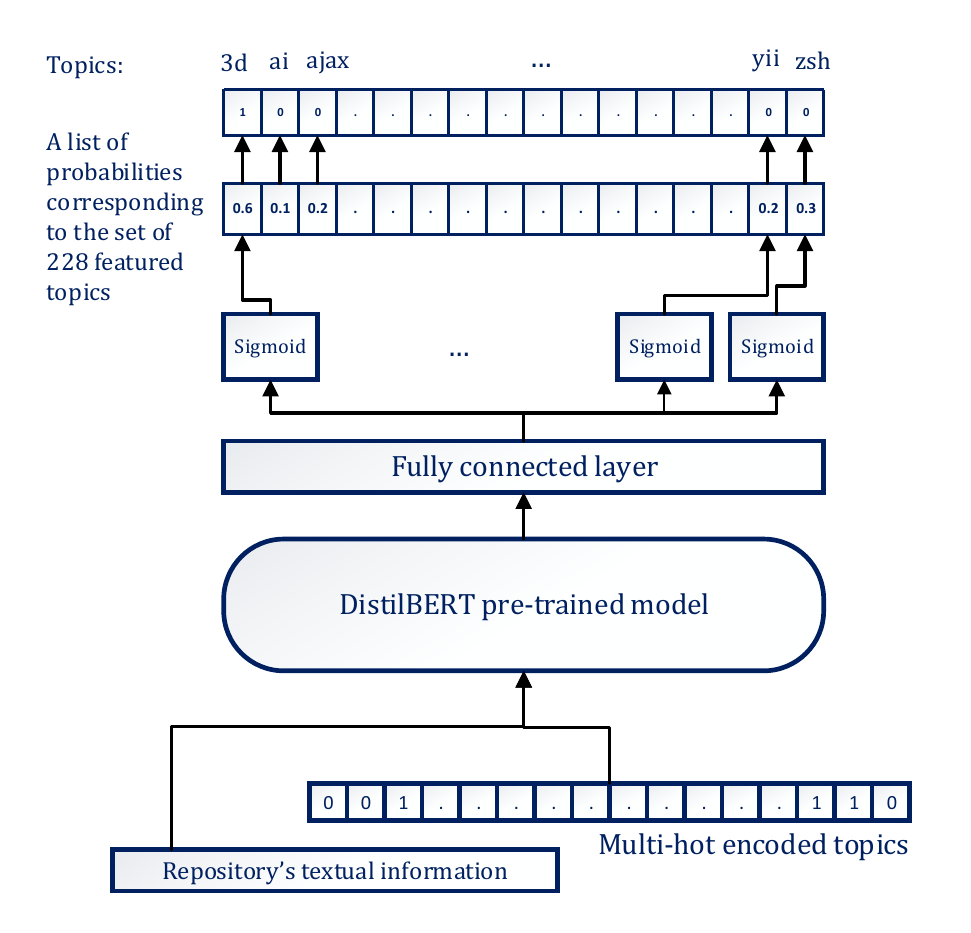}
    \caption{Fine-tuning DistilBERT for multi-label classification}
    \label{fig:fine-tune}
\end{figure}

Recently, Transformers and the BERT model 
have significantly impacted the NLP domain.
This is because the pre-trained BERT model can be fine-tuned 
with just one additional output layer to create state-of-the-art models 
for a wide range of NLP tasks (in our case multi-label classification), 
without major task-specific architecture modifications. 
Therefore, we exploit DistilBERT, 
a successful variant of BERT in our approach.
We add a multi-label classification layer on top of this model
and fine-tune it on our dataset.
Figure \ref{fig:fine-tune} depicts the architecture of our model.

\subsubsection{Handling Imbalanced Data} 
As shown in Section \ref{sec:preprocess_topics}, 
the distribution of topics in our dataset is 
very unbalanced (long-tailed distribution).
That is, most of the repositories are assigned with a very few number of topics 
while many other topics are used less frequently (have less support).
In such cases, the classifier can be biased 
toward predicting more frequent topics more often, 
hence increasing precision 
and decreasing recall of the least-frequent topics.
Therefore, we need to assign more importance 
to certain topics and define more penalties for their misclassification. 
To this end, we define a vector 
containing the weights corresponding to our topics 
in the fit method of our classifiers. 
It is a list of weights with the same length as the number of topics. 
We populate this list with a dictionary of ${topic: weight}$. 
Weight for topic $t_i$ is equal to 
the ratio of the total number of repositories denoted as $N$
to the frequency of a topic ($frequency_{t_i}$) as shown in Equation \ref{eq:class_weights}. 
Thus, less-frequent topics will have higher weights 
while calculating loss functions.
Therefore, the model learns to better predict them.
\begin{equation}
\label{eq:class_weights}
weight_{t_i} = \dfrac{N}{frequency_{t_i}}
\end{equation}

\section{Experimental Design}\label{sec:eval}
In this section, we present our experimental setting.

\subsection{Dataset and Models}\label{sec:dataset}
We divided our preprocessed dataset of GitHub repositories (Section \ref{sec:data})
to three subsets of training, validation, and testing datasets. 
We first split the data into train and test sets with ratios of $80\%$, and $20\%$, respectively. 
Then we split the train set to two subsets 
to have a validation set as well (with ratios $90\%$ to $10\%$). 
We have about $152$K repositories, with $228$ selected featured topics.
Input data consists of projects' names, descriptions, READMEs, wiki pages, and file names concatenated together.

To train traditional classifiers, 
we use the \textit{Sci-kit Learn}\footnote{https://scikit-learn.org} library. 
We exploit its \textit{OneVsRestClassifier} feature 
for some of our traditional models such as NB and LR.
Furthermore, we use the \textit{HuggingFace} \footnote{https://huggingface.co} 
and the \textit{SimpleTransformers}\footnote{https://gitbub.com/ThilinaRajapakse/simpletransformers} libraries 
for the implementation of our DistilBERT-based classifier. 
We set the learning rate to $3e-5$, 
the number of epochs to $9$, 
the maximum input length to $512$ and the batch size to $4$.
We set the maximum number of features for to $20K$ and $1K$ 
for TF-IDF and Doc2Vec embeddings.
Higher numbers would result in overfitted models 
and/or the training time would increase greatly.
We also set the minimum frequency count for Doc2Vec to $10$ 
and the ngram range to $(1, 2)$ for TF-IDF. 
As for the FastText, We first optimize it 
by setting the Automatic tuning duration to $20$ hours.
The best parameters retrieved for our data are 
the learning rate of $1.08$, 
the minimum frequency count of $1$, and the ngram size of $3$.
We set the remaining parameters to default values.
Our experiments are conducted on a server
equipped with two GeForce RTX 2080 GPUs,
an AMD Ryzen Threadripper 1920X CPU with $12$ core processors, and $64$G RAM.

Baseline models here are the Di Sipio et al's \mbox{\cite{di2020multinomial}} approach 
and variations of the Naive Bayes algorithm, namely MNB and GNB.
We choose the latter two 
because the core algorithm in the baseline \mbox{\cite{di2020multinomial}} is an MNB.
Furthermore, these techniques lack balancing 
while our proposed models use balancing techniques.
Di Sipio et al. \mbox{\cite{di2020multinomial}},
first extracts a balanced subset of the training dataset, 
by taking only $100$ sample repositories 
for each of their selected featured topics.
It then proceeds to train an MNB on this data.
In the prediction phase, 
the authors use a source code analysis tool, \textit{GuessLang},
to predict the programming language of each repository separately.
In the end, they take $n-1$ topics predicted by their classifier 
and concatenate it with the programming language topic extracted from the GuessLang 
and generate their $top-n$ recommendation list.

\subsection{Evaluation Metrics}\label{sec:eval_metric}
To evaluate our methods,
we use standard evaluation metrics applied 
in both recommendation systems and multi-label classification scenarios
such as Recall, Precision, F1 measure, Success Rate, and LRAP
to address different aspects of our model \cite{izadi2014unifying,schapire2000boostexter}.
The evaluation metrics used in our study are as follows.
\begin{itemize}
    \item \textbf{Recall, Precision, and F1 measure}: 
    These are the most commonly used metrics 
    in assessing a recommender system's performance in the top-$n$ suggested topics \cite{jalili2018evaluating}. 
    Precision is the ratio $\frac{tp}{tp + fp}$ 
    where $tp$ is the number of true positives 
    and $fp$ the number of false positives. 
    Thus, $P@n$ for a repository is the percentage of 
    correctly predicted topics among the top-$n$ recommended topics for that repository.
    Similarly, Recall is the ratio $\frac{tp}{tp + fn}$ 
    where $fn$ is the number of false negatives. 
    Thus, $R@n$ for a repository is the percentage of 
    correctly predicted topics among the topics that are actually assigned to that repository.
    $F1$ measure, as expected, is the harmonic mean of the previous two
    and is calculated as $ \frac{2 \times P \times R}{P + R}$. 
    We report these metrics for $top-n$ recommendation lists.
    Moreover, we show how much these metrics are affected 
    by changing the size of the recommendation list.
    \item \textbf{Success Rate}: 
    We denote success rate for different top-$n$ recommendation lists 
    as $S@n$ and report $S@1$ and $S@5$. S@1 measures 
    whether the most probable predicted topic for each repository, 
    is correctly predicted. S@5 measures whether 
    there is at least one correct suggestion among the top-five recommendations.
    \item \textbf{LRAP}: 
    This metric is used for multi-label classification problems, 
    where the aim is to assign better ranks to the topics 
    truly associated with each repository \cite{schapire2000boostexter}.
    That is for each ground truth topic, LRAP evaluates 
    what fraction of higher-ranked topics were true topics. 
    LRAP is a threshold-independent metric which scores between $0$ and $1$, with $1$ being the best value.
    Equation \ref{eq:lrap}, calculates LRAP. 
    Given a binary indicator matrix of the ground truth topics 
    and the score associated with each topic, the average precision is defined as
    \begin{equation}
    \label{eq:lrap}
        LRAP(y, \hat{f}) = \frac{1}{n_{\text{repositories}}}
        \sum_{i=0}^{n_{\text{repositories}} - 1} \frac{1}{||y_i||_0}
         \sum_{j:y_{ij} = 1} \frac{|\mathcal{L}_{ij}|}{\text{rank}_{ij}}
    \end{equation}
    where 
    $\mathcal{L}_{ij} = \left\{k: y_{ik} = 1, \hat{f}_{ik} \geq \hat{f}_{ij} \right\}$, 
    $\text{rank}_{ij} = \left|\left\{k: \hat{f}_{ik} \geq \hat{f}_{ij} \right\}\right|$, 
    $|\cdot|$ computes the cardinality of the set 
    that is the number of elements in the set, 
    and $||\cdot||_0$
    is the $\ell_0$ ``norm''. 
\end{itemize}

\subsection{User Study to Evaluate Recommendation Lists}
We designed a questionnaire to assess the quality of 
our recommended topics from users' perspectives.
We randomly selected $100$ repositories 
and included recommended topics 
(1) by our approach (LR with TF-IDF embeddings), 
(2) by the baseline approach (Di Sipio et al. \cite{di2020multinomial})
and (3) the set of the original featured topics. 
We present these sets of recommended topics to the participants
as outputs of three anonymous methods 
to prevent biasing them.
We asked the participants to rate 
the three recommendation lists for each repository
based on their correctness and completeness.
That is for each repository they answer the following questions:

\textbf{Correctness}: how many correct topics are included in each recommendation list,

\textbf{Completeness}: compare and rank the methods for each repository 
based on the completeness of the correct recommendations.

As this would require a long questionnaire and assessing all samples could
jeopardize the accuracy of evaluations, 
we randomly assigned the sample repositories
to the participants 
and made sure to cover each of the $100$ repositories at least 
by $5$ participants.
To provide better context, 
we also include the content of the README file of repositories for the users.

\section{Results}\label{sec:results}
In this section, we present the results of our experiments and discuss them.
We first review the results of the proposed multi-label classification models and compare them with the baselines.
Then, we present the results of the user study to assess the results from the participants' perspective.
Next, we analyze the results per topic and assess the quality of recommendations.
Finally, using the data ablation study, we address our last research question.

\subsection{RQ2: Recommendation Accuracy}
To answer \textbf{RQ2}, 
we present the results of both the baselines and the proposed models 
based on our evaluation metrics. 
We set $n = (1, 5)$, 
and report the results for $S@n$, $R@n$, $P@n$, and $F@n$ in Table \ref{tab:results_topk}. 
As shown by the results, we outperform the baselines 
by large margins regarding all evaluation metrics.
In other words, we improve the baseline \cite{di2020multinomial} 
by $29\%$,  $59\%$, $65\%$, $63\%$ and $46\%$ 
in terms of $S@5$, $R@5$, $P@5$, $F1@5$ and $LRAP$, respectively.
Among our proposed models, 
the LR classifier with TF-IDF embeddings and the DistilBERT-based classifier 
achieve similar results and both outperform all other models.
\begin{table*}[ht]
\caption{Evaluation results}
\centering
\begin{tabular}{ccc|ccc|c|cc}\toprule
    & \multicolumn{6}{c}{Evaluation Metrics}\\
    \cmidrule{2-7}
    \multicolumn{1}{l}{\textbf{Baseline \newline models}} 
                                     & S@1 & S@5 & R@5 & P@5 & F1@5  & LRAP & T(t) & T(p)\\\midrule
    \multicolumn{1}{l}{Di Sipio et al. \cite{di2020multinomial}} 
                                     & 0.465 & 0.750 & 0.561 & 0.210 & 0.289  & 0.553 & 20s & 93s\\%
    \multicolumn{1}{l}{MNB, TF-IDF}  & 0.581 & 0.833 & 0.659 & 0.253 & 0.346  & 0.569 & 3m & 0.5ms\\%
    \multicolumn{1}{l}{GNB, D2V}     & 0.604 & 0.901 & 0.753 & 0.287 & 0.393  & 0.619 & 30m & 0.6ms\\%

    \midrule
    \multicolumn{1}{l}{\textbf{Proposed \newline models}} 
                                    & S@1 & S@5 & R@5 & P@5 & F1@5  & LRAP & T(t) & T(p)\\\midrule
    \multicolumn{1}{l}{FastText}    & 0.783 & 0.958 & 0.855 & 0.330 & 0.450  & 0.772 & 25m 
                                    & 0.4ms\\
    \multicolumn{1}{l}{LR, D2V}     & 0.624 & 0.931 & 0.795 & 0.302 & 0.415  & 0.662 & 29h & 0.3ms\\%
    \multicolumn{1}{l}{LR, TF-IDF}  & \textbf{0.806} & \textbf{0.971} & \textbf{0.890} & \textbf{0.346} &                                     \textbf{0.470} & \textbf{0.805} & 30m & 0.4ms\\%
    \multicolumn{1}{l}{DistilBERT}  & \textbf{0.792} & \textbf{0.969} & \textbf{0.884} & \textbf{0.343} &                                     \textbf{0.469} & \textbf{0.796} & 10.5h & 5ms\\%
    \bottomrule
\label{tab:results_topk}
\end{tabular}
\end{table*}

Another aspect of these models' performance 
is the time it takes to train them and predict topics.
Table \ref{tab:results_topk} presents the training time 
for each model as $T(t)$ and the prediction time of a complete set of topics
for a repository as $T(p)$. 
To predict the prediction time, 
we calculate the prediction time of $1000$ sample recommendation lists 
for each model and report the average time per list.
The values are in millisecond, minutes, and hours.
Note that prediction time of the baseline \cite{di2020multinomial}
is significantly larger than our models.
This unnecessary delay is caused due to using the GuessLang tool 
for predicting programming language topics for repositories.
Although the training time is a one-time expense,
prediction time can be a key factor when choosing the best models.

\begin{figure}[h!]
    \centering
    \subfigure[]
    {\includegraphics[width=0.48\linewidth]{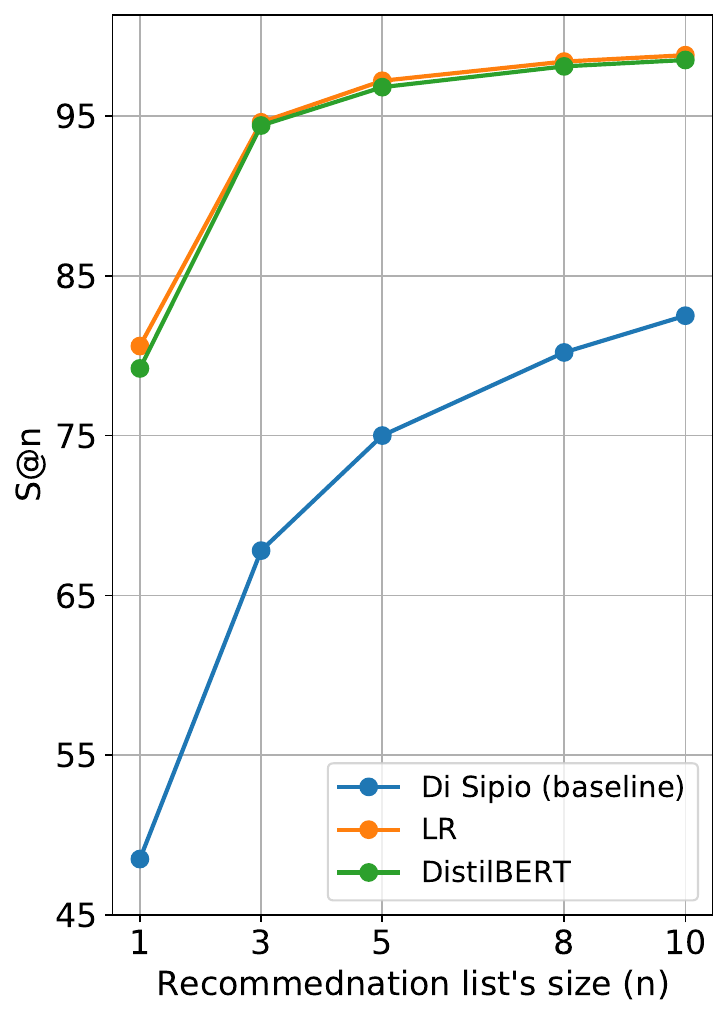}}
        \label{fig:s_r}
    \subfigure[]
        {\includegraphics[width=0.48\linewidth]{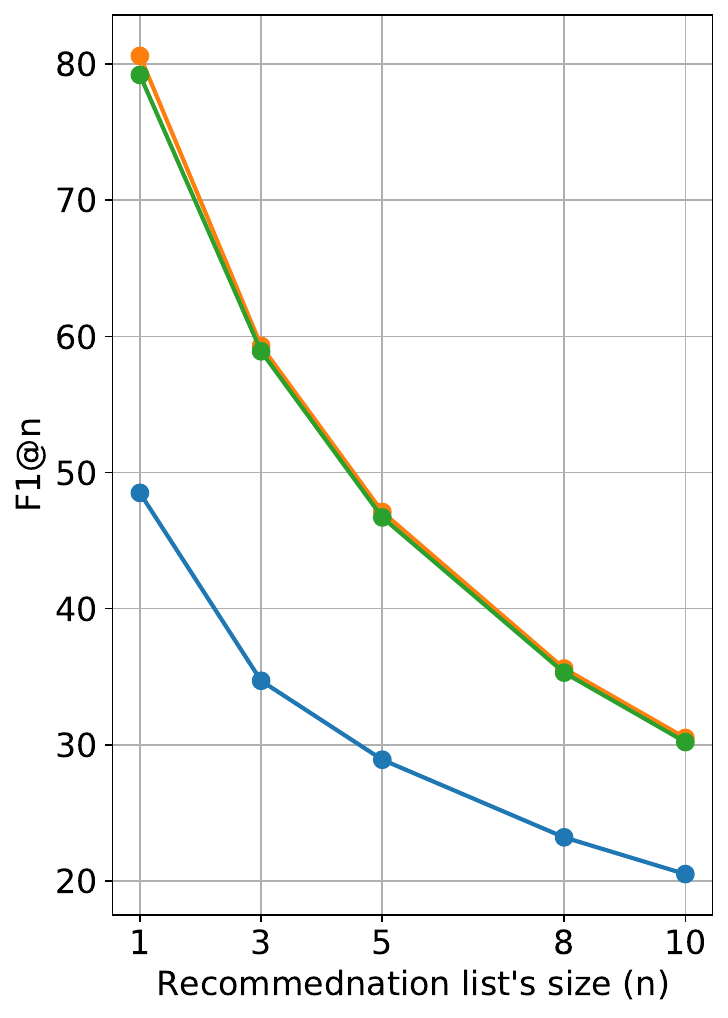}}
            \label{fig:f_r}
    \caption{Comparing results for different recommendation sizes.} 
    \label{fig:sizes}
\end{figure}

Moreover, we vary the size of recommendation lists 
and analyze their impact on the results.
We set the parameter $n$ (size) equal to $1$, $3$, $5$, $8$, and $10$, respectively,
and report the outcome in \mbox{Figure \ref{fig:sizes}}. 
As expected, as the size of recommendation list increases, so does the $S@n$.
However, while $R@n$ goes up, $P@n$ goes down and thus the $F1@n$ decreases.
Note that both LR and DistilBERT-based classifier
perform very closely regarding for all recommendation sizes and metrics.

To investigate whether there is a significant difference between the
results of our proposed approach and the baseline, 
we followed the guideline and
the tool provided by Herbold \cite{herbold2020autorank}.
We conducted a statistical analysis for three approaches 
of Di Sipio et al. \cite{di2020multinomial}, LR and DistilBERT-based classifiers
and used $30280$ paired samples.
We reject the null hypothesis 
that the population is normal for the three populations 
generated by these approaches. 
Because we have more than two populations 
and due to the fact that they are not normal,
we use the non-parametric \textit{Friedman} test to investigate the differences between the
median values of the populations \cite{friedman1940comparison}. 
We employed the \textit{post-hoc Nemenyi} test to determine
which of the aforementioned differences 
are statistically significant \cite{nemenyi1962distribution}. 
The Nemenyi test uses critical distance (CD) 
to evaluate which one is significant. 
If the difference is greater than CD, 
then the two approaches are statistically significantly different.
We reject the null hypothesis of the Friedman test that 
there is no difference in the central tendency of the populations. 
Therefore, we assume that there is a statistically significant difference 
between the median values of the populations.
Based on the post-hoc Nemenyi test, we assume that 
there are no significant differences within the following groups: 
LR and DistilBERT-based classifier. 
All other differences are significant.

\begin{figure}
    \centering
    \includegraphics[width=0.8\linewidth]{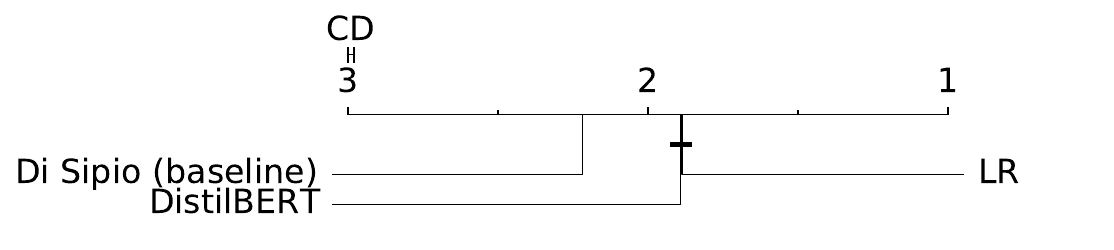}
    \caption{The results of hypothesis testing for F1@5 measure}
    \label{fig:autorank_size}
\end{figure}

Figure \ref{fig:autorank_size} depicts 
the results of hypothesis testing for F1@5 measure.
The Friedman test rejects the null hypothesis 
that there is no difference between median values of the approaches. 
Consequently, we accept the alternative hypothesis 
that there is a difference between the approaches. 
Based on the Figure \ref{fig:autorank_size} and the post-hoc Nemenyi test, 
we cannot say that there are significant
differences within the following approaches: (LR and DistilBERT).
All of the other differences are statistically significant.

\subsection{RQ3: Results of the User Study}
\begin{figure}[h!]
\caption{User study's results}
    \subfigure[Number of Correct topics (1 to 5)]
    {\includegraphics[width=0.48\linewidth]{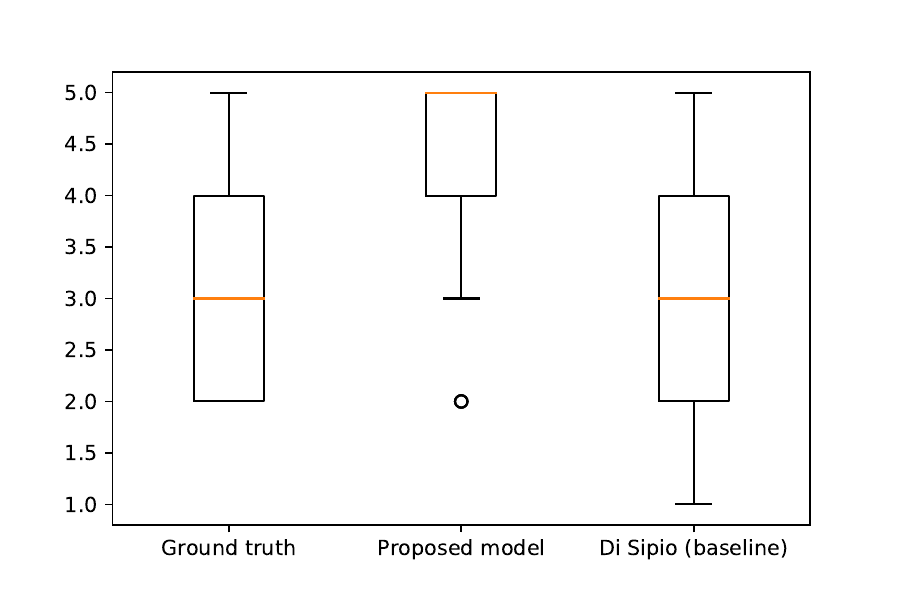}}
        \label{fig:corr}
    \subfigure[Completeness Rank (1 to 3)]
    {\includegraphics[width=0.48\linewidth]{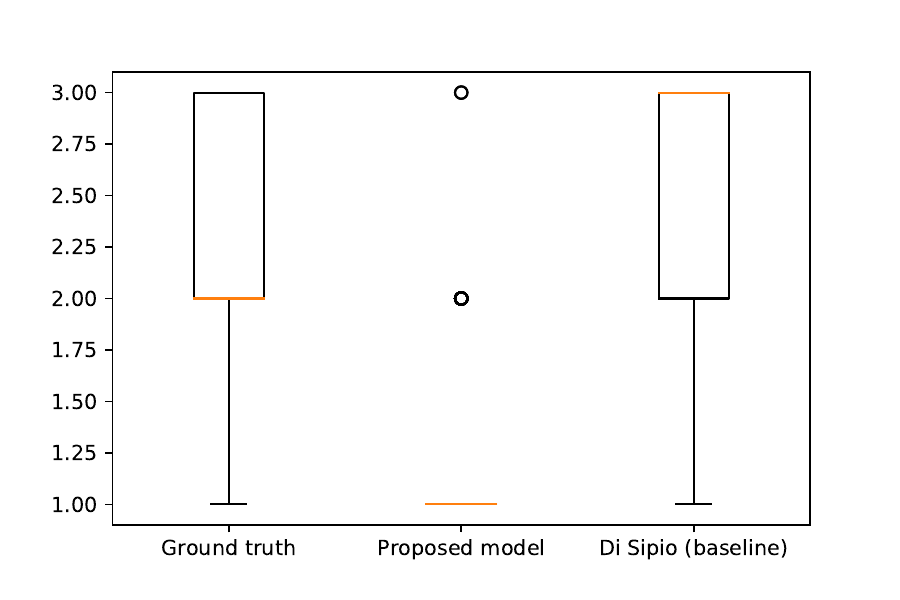}}
        \label{fig:com}
    \label{fig:userstudy}
\end{figure}

Figure \ref{fig:userstudy} shows two groups of 
BoxPlots comparing the correctness and completeness of 
the recommended topics by our three methods included in the user study.
With regard to the \textit{correctness} of the suggestions, 
the median and average correct topics of our model are $5$ and $4.48$ out of $5$ recommended topics.
While the median and average of the baseline approach, 
Di Sipio et al. \cite{di2020multinomial}, 
are $3$ and $3.07$ correct topics out of $5$ recommended topics.
Regarding the \textit{completeness} of the suggestions, 
the median and average rank assigned 
by the participants to our approach are $1$ and $1.2$, respectively.
This means almost in all cases, 
our approach recommends the most complete set of correct topics.
Although there are a couple of outlier cases in which 
our proposed approach is ranked second or third (Figure \ref{fig:corr}-b).
The median and average of the assigned rank for the baseline method are $3$ and $2.4$ correct topics. 
That is in most cases, participants ranked the baseline as the last approach in terms of completeness.
Note that we did not ask the participants 
to score the recommendations based on the \textit{usefulness} of individual topics. 
This is because, to the best of our knowledge, 
there is no agreement on \textit{what is a useful topic} 
in the related work yet. 
However, we asked an open question on what is \textit{a useful set of topics}. 
Specifically, we asked ``What do you consider a useful set of recommended topics"? 
In our study, participants mostly emphasized the completeness of the sets. 
For instance, one participant stated:

\textit{``More complete sets of topics make it easier 
to select suitable topics for my repositories 
because they can point out different aspects such as the goal, 
the platform it can be used on, its category, the languages, etc. 
So for me, the higher number of correct topics 
equals the usefulness of the recommended set."}

According to the results, 
we can conclude that our recommendations are also deemed useful by developers.
Moreover, our approach can recommend missing topics as well.
In fact, users indicated that our recommended topics 
often were more complete than featured topics of the repositories.
This is probably because repository owners 
sometimes forget to tag their repositories with a complete set of topics. 
Thus, some correct topics will be missing from the repository (missing topics).
However, our ML-based model has learned from the dataset 
and is able to predict more correct topics.
This also can be the reason for the low Precision score 
of the ML-based models because the ground truth 
is lacking some useful and correct topics.
As will be shown in the Data Ablation Study next section,
by mapping user-defined topics to featured topics 
we are able to extract more valuable information from the data 
and indeed increase scores of Precision and F1 measure.

Therefore, to answer RQ3, 
we conclude that our approach can successfully 
recommend accurate topics for repositories. 
Moreover, it is able to recommend more complete sets 
comparing to both the baseline's and the featured sets of topics.

\subsection{Qualitative Analysis of the Recommendations}
Table \ref{tab:results_sample_list} 
presents our model's recommended topics for a few sample repositories. 
As confirmed by the user study,
our proposed approach is not only capable of recommending correct topics 
but also it can recommend missing topics. 
For instance, the \textit{sherbold/autorank}
is a Python package for comparing paired populations.
Currently, this repository does not have any original topics, however, 
our model's top five recommendations are all correct.
The recommendations show that our model not only can detect coarse-grained features 
such as the programming language or the general category of a repository 
such as \texttt{python}, \texttt{machine-learning}, and  \texttt{algorithm}, 
but also it is able to recommended proper functionality- or goal-related topics 
such as \texttt{data-visualization} and \texttt{testing} which are more fine-grained and specific.
Below, we present a list of such specific topics (e.g., functionality-related) along their recall score as an indication of the performance of our LR-based model on these topics:\\
\texttt{3d (79\%), bioinformatics (79\%), blockchain (90\%), cli (77\%), cms(84\%), compiler (82\%), composer (73\%), computer-vision (84\%), cryptocurrency (88\%), data-structures (82\%), data-visualization (76\%), database (77\%), deep-learning (93\%), docker (88\%), emulator (87\%), game-engine (83\%), google-cloud (81\%), home-assistant (93\%), image-processing (78\%),  localization (70\%), machine-learning (84\%), monitoring (78\%), neural-network (92\%), nlp (89\%), opencv (85\%), package-manager (68\%), robotics (74\%), security (77\%), testing (74\%), virtual-reality (83\%), web-components (82\%), webextension (93\%), webpack (81\%)}, etc.
\begin{table*}[ht]
\caption{Recommendations for sample repositories}
\centering
\begin{tabular}{p{45mm}p{25mm}p{35mm}}\toprule
    Repositories                      &  Featured topics & Recommended topics (LR)\\\midrule
\textbf{sherbold/autorank} \\
A Python package to
simplify the comparison between (multiple) paired populations.
                                      & - 
                                      & \textcolor{PineGreen}{python}, 
                                      \textcolor{PineGreen}{machine-learning}, \textcolor{PineGreen}{data-visualization}, 
                                      \textcolor{PineGreen}{testing}, 
                                      \textcolor{PineGreen}{algorithm}\\\midrule 
\textbf{parrt/dtreeviz} \\
A python library for decision tree visualization and model interpretation.
                                      & -
                                      & \textcolor{PineGreen}{machine-learning}, \textcolor{PineGreen}{scikit-learn}, 
                                      \textcolor{PineGreen}{data-visualization}, 
                                      \textcolor{PineGreen}{python}, 
                                      \textcolor{PineGreen}{ai}\\\midrule 
\textbf{iterative/dvc} \\
Git for Data and Models.              & python, git, data-science, machine-learning, ai
                                      & \textcolor{PineGreen}{Python}, 
                                        \textcolor{PineGreen}{git}, 
                                       yaml,
                                       \textcolor{PineGreen}{terminal},  
                                       \textcolor{PineGreen}{machine-learning} \\\midrule 
\textbf{plotly/dash} \\
Analytical Web Apps for Python, R, Julia, and Jupyter.
 	                                  & data-visualization, react, data-science, python, bioinformatics
                                      & \textcolor{PineGreen}{data-visualization}, 
                                      \textcolor{PineGreen}{react},  
                                      \textcolor{PineGreen}{python}, 
                                      kubernetes,
                                      \textcolor{PineGreen}{ai} \\\midrule 
\textbf{pypa/pip} \\
The Python package installer
 	                                  & python, pip
                                      & \textcolor{PineGreen}{python}, 
                                      \textcolor{PineGreen}{pip},  
                                      \textcolor{PineGreen}{package-manager}, 
                                      \textcolor{PineGreen}{dependency-management}, 
                                      yaml \\
\bottomrule
\label{tab:results_sample_list}
\end{tabular}
\end{table*}

In Table \ref{tab:results_bestClasses}, 
we presents the results based on different topics. 
About $100$ topics have Recall and Precision scores 
higher than $80\%$ and $50\%$, respectively. 
Furthermore, only six topics out of $228$ topics have Recall scores lower than $50\%$. 
Thus, in the following we will investigate cases for which the model reports low Precision.
We divide these topics into two groups: 
(1) topics assigned to a low number of repositories (weakly-supported topics), and
(2) topics assigned to a high number of repositories (strongly-supported topics).
In the first row, we report $36$ topics of the first group, such as
\texttt{phpunit}, \texttt{code-review}, \texttt{less}, 
\texttt{storybook}, \texttt{code-quality}, and \texttt{package-manager} 
that are assigned to repositories less than $80$ times in our data. 
Note that we employ used balancing techniques in our models, 
which help recommend less-frequent and specific topics correctly as much as possible. 
However, some of these topics seem to convey concepts used 
in general cases such as 
\texttt{operating-system}, \texttt{privacy}, \texttt{npm}, \texttt{mobile}, and \texttt{frontend}.
Therefore, we believe augmenting the dataset 
with more sample repositories tagged by these topics
can boost the performance of our classifiers.
Thus, when collecting new data points, 
both the support number of weakly-supported topics with low precision ($80$) 
and the cutoff threshold in our dataset ($100$) should be taken into account.

In the second row, we have $12$ popular topics, 
namely \texttt{javascript}, \texttt{library}, \texttt{api}, 
\texttt{framework}, \texttt{nodejs}, \texttt{server},\texttt{linux}, \texttt{html}, 
\texttt{c}, \texttt{windows}, \texttt{rest-api}, and \texttt{shell}
for which the model achieves good recall scores (higher than $70\%$), 
but low recall precision scores ($20\%$ to $40\%$). 
The numbers of repositories tagged with these topics ranges from $300$ to $2900$.
As the number of sample repositories seem to be sufficient, 
the low precision of the model can be due to several reasons.
Upon investigation, we found out some users 
often forget to assign general-purpose topics. 
That is the programming language of a repository can be indeed JavaScript 
or the operating system can be Linux or Windows. 
But users neglect tagging their repositories with these general purpose topics, 
hence the ground truths will be missing these correct topics.
Then, when the trained model predicts these missing topics correctly, 
it will be penalized since they are missing from ground truth.
Subsequently, this will result in low Precision scores for these topics.
Second, some of these topics such as \texttt{api}, \texttt{framework} and \texttt{library}
have extensive broadness, popularity, and subjectiveness.  
For instance, users often mix the above-mentioned topics 
and use them interchangeably or subjectively. 
And any machine learning model is only as good as the data it is provided with.

\begin{table*}[ht]
\caption{Performance based on topics}
\centering
\begin{tabular}{p{30mm}p{80mm}}\toprule
                 & Featured Topics \\\midrule
    Low precision,\newline and weakly-supported  
    & \texttt{operating-system}, \texttt{p2p}, \texttt{privacy}, \texttt{neovim}, \texttt{eslint}, \texttt{yaml}, \texttt{hacktoberfest}, \texttt{aurelia}, \texttt{csv}, \texttt{web-components}, \texttt{gulp}, \texttt{maven}, \texttt{styled-components}, \texttt{homebrew}, \texttt{mongoose}, \texttt{nuget}, \texttt{firefox-extension}, \texttt{threejs}, \texttt{localization}, \texttt{wpf}, \texttt{scikit-learn}, \texttt{pip}, \texttt{webextension}, \texttt{virtual-reality}, \texttt{github-api}, \texttt{ajax}, \texttt{archlinux}, \texttt{nosql}, \texttt{vanilla-js}, \texttt{package-manager}, \texttt{less}, \texttt{storybook}, \texttt{code-quality}, \texttt{dependency-management}, \texttt{code-review}, \texttt{phpunit}
\\\midrule
    Low precision,\newline but strongly-supported  
    & \texttt{javascript}, \texttt{library}, \texttt{api}, 
    \texttt{framework}, \texttt{nodejs}, \texttt{server},\texttt{linux}, \texttt{html}, 
    \texttt{c}, \texttt{windows}, \texttt{rest-api}, \texttt{shell} \\
\bottomrule
\label{tab:results_bestClasses}
\end{tabular}
\end{table*}

\subsection{RQ4: Data Ablation Study}\label{sec:data_ablation}
To answer \textbf{RQ4}, we train our proposed models
using different types of repository information 
(i.e., description, README, wiki pages, and file names) as the input. 
According to the results (Table \ref{tab:results_diff_inputs}), 
as a single input, wiki pages have the least valuable information. 
This is probably because only a small number of repositories 
(about $10\%$) contained wiki pages
and it appears these pages are often missing from repositories. 
On the other hand, among single source inputs, READMEs provide better results. 
This is probably because READMEs are the main source 
for providing information about a repositories' goals and characteristics.
Thus, they have an advantage compared 
to other sources regarding both the quality and quantity of tokens.
Consequently, READMEs are enabled to contribute more to training. 
While READMEs are essential for training models,
Therefore, To answer RQ4, adding more sources of information 
such as descriptions and file names 
indeed helps boost the models' performance.
Furthermore, these information complement each other 
in case a repository does not have a description, 
README or adequate number of files at the same time.
\begin{table*}[ht]
\caption{Evaluation results based on different types of input}
\centering
\begin{tabular}{cccccc}\toprule
                        & \multicolumn{4}{c}{Evaluation Metrics}&\\
    \cmidrule{2-5}
    LR, TFIDF          & R@5 & P@5 & F1@5 & LRAP & support\\
   \midrule
    Wiki pages          & 45.6\% & 18.8\% & 25.1\% & 39.0\% & 3.5K\\
    File names          & 71.4\% & 27.3\% & 37.4\% & 62.2\% & 30K\\
    Description         & 72.2\% & 27.8\% & 38.0\% & 65.7\% & 30K\\
    README              & 84.3\% & 32.6\% & 44.5\% & 75.2\% & 30K\\
    All but file names  & 86.7\% & 33.6\% & 45.9\% & 78.1\% & 30K\\
    ALL                 & \textbf{89.0\%} & \textbf{34.6\%} & \textbf{47.0\%} & \textbf{80.5\%} & 30K\\
    \midrule
    DistilBERT          & R@5 & P@5 & LRAP & F1@5 & support\\
    \midrule
    Wiki pages          & 30.4\% & 12.6\% & 16.8\% & 26.2\%& 3.5K\\
    File names          & 67.1\% & 25.4\% & 34.8\% & 58.8\% & 30K\\
    Description         & 71.0\% & 27.2\% & 37.3\% & 64.2\% & 30K\\
    README              & 84.2\% & 32.5\% & 44.4\% & 74.9\% & 30K\\
    All but file names  & 86.6\% & 33.5\% & 45.8\% & 77.8\% & 30K\\
    ALL                 & \textbf{88.4\%} & \textbf{34.3\%} & \textbf{46.9\%} & \textbf{79.6\%} & 30K\\
    \bottomrule
\label{tab:results_diff_inputs}
\end{tabular}
\end{table*}

\subsubsection{Different Number of Topics}
\begin{figure}[h!]
    \centering
    \subfigure[]
        {\includegraphics[width=0.48\linewidth]{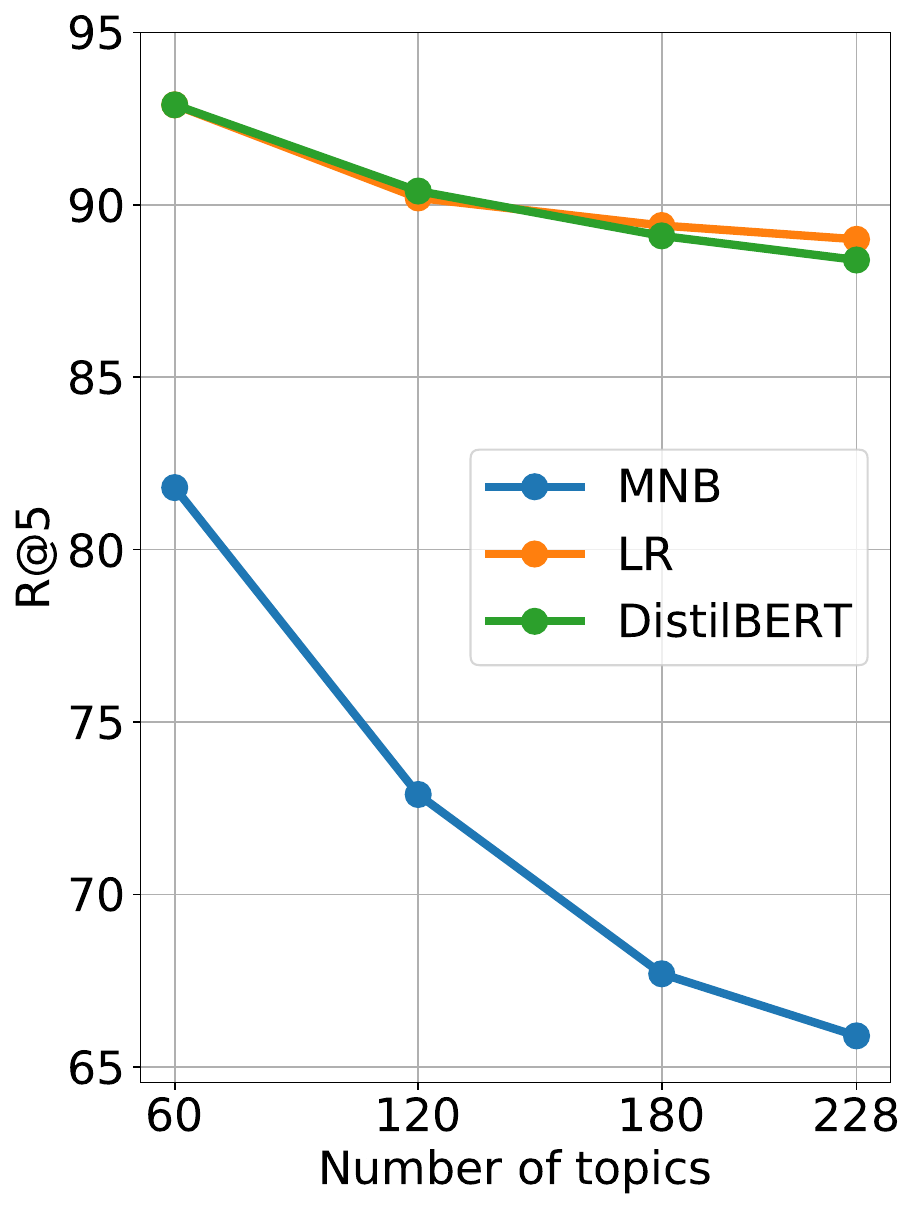}}
    \subfigure[]
        {\includegraphics[width=0.48\linewidth]{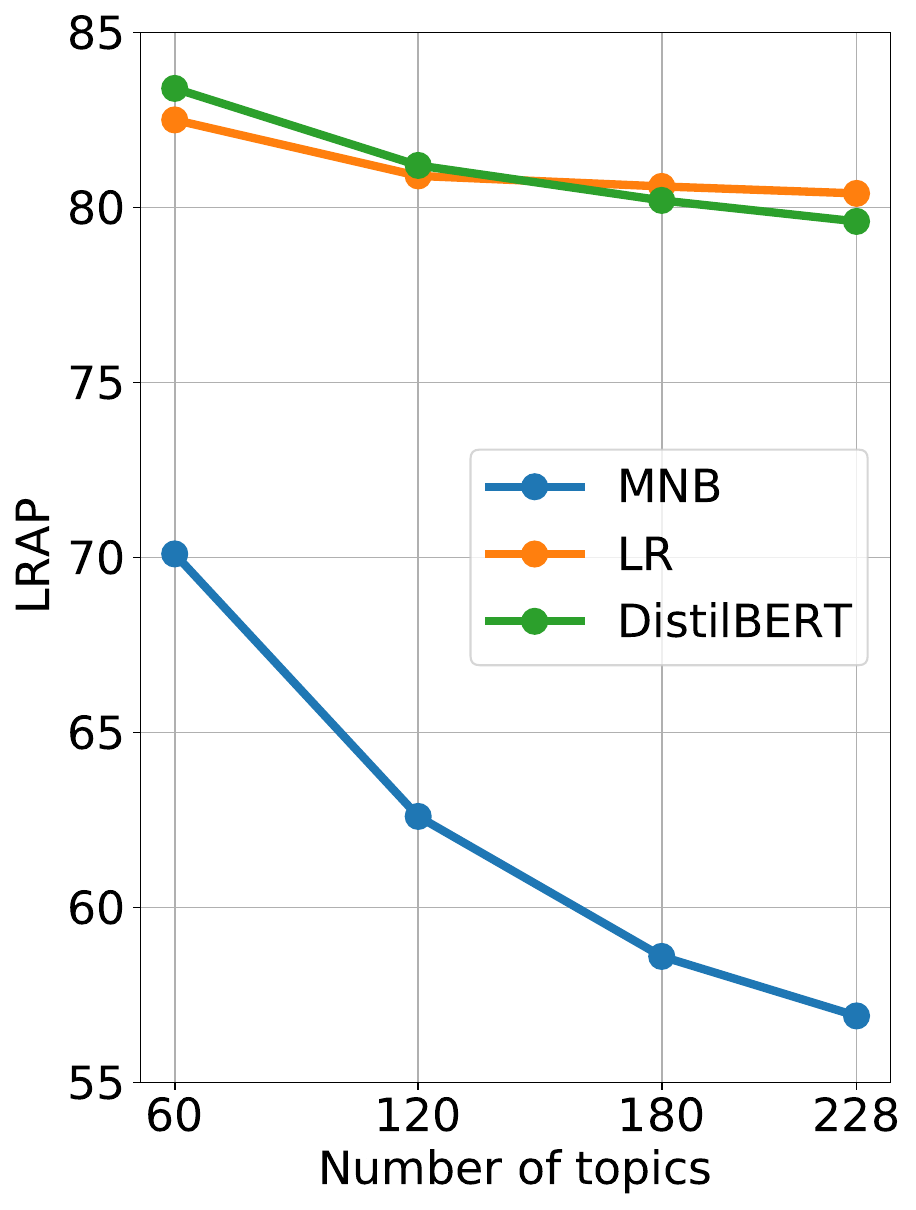}}
    \caption{Comparing results for different number of topics.} 
    \label{fig:num_labels_at_n}
\end{figure}

We also investigate whether 
there is a relationship between the performance of different models 
and number of topics they are trained on.
We train several models 
on the most frequent $60$, $120$, $180$, and $228$ featured topics, respectively.
Figure \ref{fig:num_labels_at_n} depicts the results of this experiment.
The interesting insight here 
is that both our proposed models (LR and DistilBERT-based classifier)
start from the same score for each metric 
and are almost always overlapping for all number of topics.
This is shown in our qualitative analysis of the results as well 
(negligible difference between these two models).
On the other hand, the MNB classifier (baseline) both 
starts from much lower scores and decreases faster as well.

\subsubsection{Training with Separate Inputs}
Here we report the results of training the models with separate input data. 
Repository's description, README and wiki pages are consisted of sentences, 
thus they are inherently sequential.
On the other hand, file names do not have any order. 
Therefore, we separate 
(1) descriptions, README files and wiki pages from 
(2) project names and source file names 
and feed them separately to the models.
For TF-IDF embeddings, 
we set the maximum number of features to $18$K and $2$K for textual data and file names, respectively.
This is because most of the input of our repositories 
consists of textual information (descriptions, README files and wiki pages).
In the same manner, 
we set the maximum number of features to $800$ and $200$ for Doc2Vec vectors.
Then we concatenated these vectors and fed them to the models.
Table \ref{tab:results_two_inputs} shows the results of this experiment.
Interestingly, baseline models behave differently.
For instance, MNB improves, while GNB under-performs the previous case.
However, our proposed model's performance is not affected significantly.
Therefore, one should take into account these differences 
while choosing the models and their settings.
\begin{table*}[ht]
\caption{Evaluation results based on separate vs. single input data}
\centering
\begin{tabular}{cccccc}\toprule
    &  & \multicolumn{4}{c}{Evaluation Metrics}\\    
    \cmidrule{3-6}
    Models& Options             & R@5 & P@5 & F1@5 & LRAP\\
    \cmidrule{1-6}
    MNB, TF-IDF & Separate inputs& 71.0\% & 27.2\% & 37.2\% & 62.0\%\\%
    & Single inputs              & 65.9\% & 25.3\% & 34.6\% & 56.9\%\\%
    \cmidrule{1-6}    
    GNB, D2V & Separate inputs  & 58.0\% & 22.0\% & 30.2\% & 41.9\%\\
    & Single inputs             & 75.3\% & 28.7\% & 39.3\% & 61.9\%\\%
    \cmidrule{1-6}
    LR, TF-IDF & Separate inputs& \textbf{88.0\%} & \textbf{34.1\%} & \textbf{46.6\%} & \textbf{79.4\%}\\%
    & Single inputs             & \textbf{89.0\%} & \textbf{34.6\%} & \textbf{47.0\%} &\textbf{ 80.5\%}\\%
    \cmidrule{1-6}
    LR D2V & Separate inputs    & \textbf{79.7\%} & \textbf{30.3\%} & \textbf{41.6\%} & \textbf{67.0\%}\\%
    & Unit inputs               & \textbf{79.5\%} & \textbf{30.2\%} & \textbf{41.5\%} & \textbf{66.2\%}\\%
    \bottomrule
\label{tab:results_two_inputs}
\end{tabular}
\end{table*}

\subsubsection{Training before and after Topic Mapping}
Table \ref{tab:results_without_subtopics} compares several models trained 
on only featured topics versus all mapped topics 
(subtopics mapped to their corresponding featured topics).
Our results indicate that adding more featured topics through mapping sub-topics 
in all cases, improves the results in terms of Precision and F1 measure.
However, it is expected that there would be a slight decrease in the Recall score 
due to the increase in the number of true topics in the dataset.
\begin{table*}[ht]
\caption{Evaluation results before vs. after topic mapping}
\centering
\begin{tabular}{ccccc}\toprule
           &                          & \multicolumn{3}{c}{Evaluation Metrics}\\
    \cmidrule{3-5}
    Models &   Options                & R@5 & P@5 & F1@5\\
    \midrule
    MNB, TF-IDF & Before & \textbf{66.2\%} & 21.7\% & 31.1\%\\ 
    & After       & 65.9\% & \textbf{25.3\%} & \textbf{34.6\%}\\
    \midrule
    LR, TF-IDF & Before & \textbf{90.9\%}  & 30.2\%  & 43.1\%\\
    & After       & 89.0\% & \textbf{34.6\%} & \textbf{47.0\%}\\
    \midrule
    DistilBERT & Before & \textbf{89.8\%} & 29.7\% & 42.5\%\\
    & After       & 88.4\% & \textbf{34.3\%} & \textbf{46.9\%}\\
    \bottomrule
\label{tab:results_without_subtopics}
\end{tabular}
\end{table*}

\section{Practical Implications and Future Work}
\begin{figure}
    \centering
    \includegraphics[width=\textwidth]{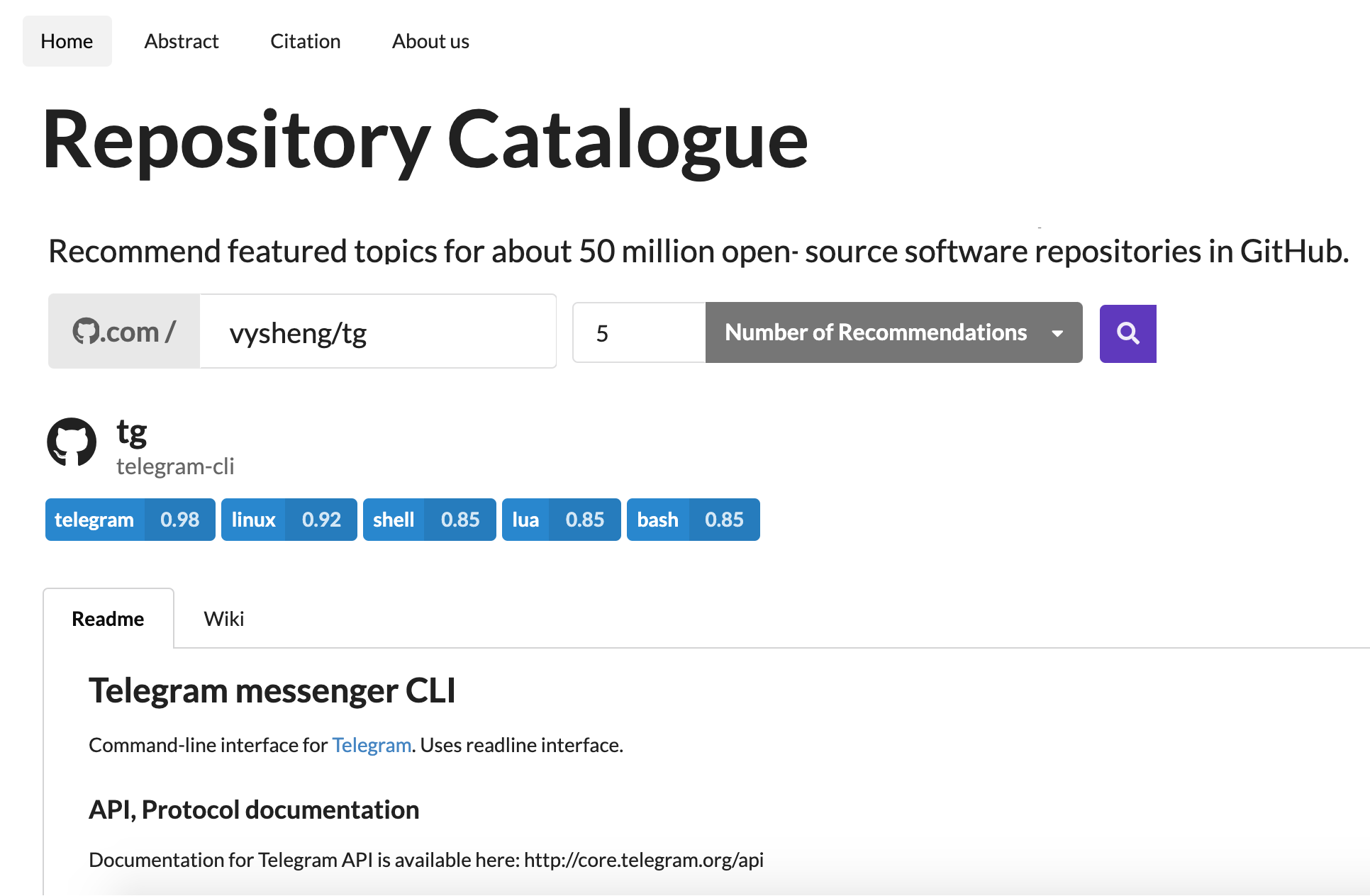}
    \caption{A screenshot of our online tool, Repologue}
    \label{fig:website}
\end{figure}

One of the major challenges 
in management of software repositories is 
to provide an efficient organization of software projects such that 
users will be able to easily navigate through the projects 
and search for their target repositories. 
Our research can be the grounding step towards a solution for this problem. 
The direct value of topic recommenders is to 
assign various type of topics (both specific and generic) to repositories 
and maintain the size and quality of the topics set. 
In this work, we have tried to tackle this problem.
Figure \ref{fig:website} 
presents a screenshot of our online tool,
\textit{Repologue}\footnote{\url{https://www.repologue.com/}}.
Our tool recommends the most related featured topics 
for any given public repository on GitHub.
Users enter the name of the target repository 
and ask for recommendations.
Repologue will first retrieve both textual information 
and file names of the queried repository. 
Then using our trained LR model, 
it will recommend the top topics sorted based on their corresponding probabilities to the user.
Suppose a developer is coding 
using \textit{Django} framework and \textit{Python} programming language.
She is looking for a library on \textit{testing} 
that can be easily installed using \textit{pip}, her package installer.
A library such as \mbox{\href{https://github.com/jazzband/django-nose}{django-nose}} is a suitable candidate.
However, its owner has not assigned any topic to this repository and users may not find it easily.
Our tool recommends the following topics for this repository; 
\texttt{python}, \texttt{django}, \texttt{testing}, and \texttt{pip}.
Each of these topics addresses one aspect of this project.
Using our tool, owners can easily 
make their repositories more visible 
and users can find their target repositories faster.

In the next step, the set of tagged repositories can also be the input 
to a more coarse-grained classification technique for software repositories.
Such a classifier can facilitate the navigation task for users. 
In other words, the next steps to our research could be 
to analyze these topics, find the relationship between them, 
and build a taxonomy of topics. 
Then, using this taxonomy, one can identify the major classes 
existing in software repositories
and build a classification model 
for categorizing repositories in their respective domain. 
Such categorization can help organize these systems 
and users will be able to efficiently search 
and navigate through software repositories. 
Another approach could be to utilize topics as a complementary input in a search engine. 
Current search engines mainly operate based on the 
similarity of textual data in the repositories. 
Feeding these topics as a weighted input to the search engines 
can improve the search results.

\section{Related Work} \label{sec:related}
In this section, we review previous approaches to this research problem.
We organize related work in the following subgroups, 
including approaches on 
(i) predicting the topic of a software repository, 
and (ii) recommending topics for other software entities. 
\subsection{Topic Recommendation for GitHub Repositories}
In 2015, Vargas-Baldrich et al. \cite{vargas2015automated}, 
presented \textit{Sally}, a tool to generate tags for Maven-based software projects 
through analyzing their bytecode and the dependency relations among them with. 
This tool is based on an unsupervised multi-label approach. 
Unlike this approach, we have employed supervised machine-learning-based methods. 
Furthermore, our approach does not require inspecting the bytecode of programs, 
and hence, can be used for all types of repositories. 

Cai et al. \cite{cai2016greta} proposed a graph-based cross-community approach, 
\textit{GRETA}, for assigning topics to repositories. 
The authors built a tagging system for GitHub by constructing an Entity-Tag Graph 
and taking a random walk on the graph to assign tags to repositories. 
Note that this work was conducted in 2016, 
prior to the time that GitHub enabled users to assign topics to repositories, 
thus the authors focused on building the tagging system from scratch 
and use cross-community domain knowledge, 
i.e. question tags from Stack Overflow QA website.
Contrary to this work, for training our model 
we used topics assigned by GitHub developers 
who actually own these repositories 
and are well aware of their salient characteristics and core functionality. 
Furthermore, the final set of topics, i.e. the featured topics, 
are carefully selected by SE community and the GitHub official team. 
Therefore, apart from applying different methods, 
the domain knowledge, quality of topics, 
and their relevance to the repositories in our work are much accurate and relevant.

Although both works have concentrated on building a tagging system 
for exploring and finding similar software projects, 
they differ in the approach and the type of input information.

Just recently, Di Sipio et al. \cite{di2020multinomial} 
proposed using an MNB algorithm for classification of about 134 topics from GitHub. 
In each top-$k$ recommendation list for a repository, 
authors would predict $k-1$ topics using the MNB (text analysis) 
and one programming language topic using a tool called \textit{GuessLang} (source code analysis). 

Similar to our work, 
they have used featured topics for training multi-label classifiers. 
However, we perform rigorous preprocessing techniques 
on both user-defined topics and the input textual information. 
We provide and evaluate a dataset of $29$K sub-topics 
for mapping to $228$ featured topics.
Our human evaluation of this dataset has shown that
we successfully map these topics and thus, 
we are able to extract more valuable information 
out of the repositories' documentation.
Not only do we consider README files, but also we process 
and use other sources of available textual information 
such as descriptions, projects and repository names, wiki pages, 
and finally file names in the repositories.
The \textit{Data Ablation Study} confirms that each type of the information we introduce to the model improves its performance.
Furthermore, we apply more suitable supervised models and balancing techniques. 
As a result of our design choices, 
we outperform their model by a large margin 
(from $59\%$ to $65\%$ improvement in terms of $R@5$ and $P@5$).
We also perform a user study 
and assess the quality of our recommendation from users' perspectives.
Our approach outperforms the baseline in this regard as well.
Finally, we have also developed an online tool that predicts topics for given repositories.

Note that we believe since GitHub already provides the programming-language of each repository 
using a thorough code analysis approach on all its source code files, 
there is not much need for predicting only the programming-language topics using code analysis. 
However, we believe code analysis can be used for more useful goals 
such as finding the relations between topics through analyzing API calls, etc. 
For instance, 
while Linares et al. \mbox{\cite{linares2014using}} 
exploits API calls for classifying applications, 
MUDABlue \mbox{\cite{kawaguchi2006mudablue}} 
and LACT \mbox{\cite{tian2009using}} use NLP techniques for this purpose. 
The result can also be used for facilitating tasks such as repository navigation.

\subsection{Tag Recommendation in Software Information Sites}
There are several pieces of research on tag recommendation 
in software information websites such as Stack Overflow, Ask Ubuntu, Ask Different, and Super User \cite{wang2018entagrec++,wang2014tag,zhou2017scalable,xia2013tag,liu2018fasttagrec,maity2019deeptagrec}. 
Question tags have been shown to help users get answers for their questions faster \cite{wang2018entagrec++}. 
They have helped in detecting and removing duplicate questions. 
Also, it has been shown that more complete tags support developers learning 
(through easier browsing and navigation) \cite{held2012learning}.
The discussion around these tags and their usability in the SE community 
have been so fortified that 
the Stack Overflow platform has also developed a tag recommendation system of their own.

These approaches mostly employ word similarity-based and semantic  sim-\newline ilarity-based techniques.
The first approach \cite{xia2013tag} focuses on 
calculating the similarity based on the textual description. 
Xia et al. \cite{xia2013tag} proposed, \textit{TagCombine}, 
to predict tags for questions using 
a multi-label ranking method based on OneVsRest Naive Bayes classifiers. 
It also uses a similarity-based ranking component, and a tag-term based ranking component. 
However, the performance of this approach is limited by the semantic gap between questions. 
Semantic similarity-based techniques \cite{wang2018entagrec++,wang2014tag,liu2018fasttagrec} 
consider text semantic information and perform significantly better than the former approach.
Wang et al. \cite{wang2018entagrec++,wang2014tag}, 
proposed \textit{ENTAGREC} and \textit{ENTAGREC++}. 
These two use a mixture model based on LLDA which considers all tags together. 
They contains six processing components: Preprocessing Component (PC), 
Bayesian Inference Component (BIC), Frequentist Inference Component (FIC), 
User Information Component (UIC), Additional Tag Component (ATC), and Composer Component (CC). 
They link historical software objects posted by the same user together.
Liu et al. \cite{liu2018fasttagrec}, proposed \textit{FastTagRec}, 
for tag recommendation using 
a neural-network-based classification algorithm and bags of n-grams (bag-of-words with word order).

\section{Threats to the Validity}\label{sec:threats}
In this section, 
we review threats to the validity of our research findings 
based on three groups of internal, external, and construct validity \cite{feldt2010validity}.

\textbf{Internal validity} relates to the variables 
used in the approach and their effect on the outcomes.
The set of topics used in our study 
can affect the outcome of our approach. 
As mentioned before a user can generate topics in free-format text, 
thus we need an upper bound on the number of topics used for training our models. 
To mitigate this problem, 
we first carefully preprocessed all the topics available in the dataset. 
Then we used the community-curated set of featured topics
provided by the GitHub team. 
We mapped our processed sub-topics 
to their corresponding featured topics, 
and finally extracted a set of a polished, widely used set of $228$ topics.
To assess the accuracy of these mappings, 
we performed a human evaluation on a randomly selected subset of the dataset.
According to the results, the Success Rate of our mapping was $98.6\%$.
We then analyzed the failed cases and update our dataset accordingly 
to avoid misleading the models
while extracting more information from the repositories' documentation.
Another factor can be errors in our code 
or in the libraries that we have used. 
To reduce this threat, we have double-checked the source code. 
But there still could be experimental errors in the set up that we did not notice.
Therefore, we have released our code and dataset publicly, 
to enable other researchers in the community to replicate it\footnote{\url{https://github.com/MalihehIzadi/SoftwareTagRecommender}}.

\textbf{Compatibility} We have evaluated the final recommended topics 
both quantitatively and qualitatively.
As shown in previous sections, their outcomes are compatible.

\textbf{External validity} 
refers to the generalizability of the results.
To make our results as generalizable as possible, 
we have collected a large number of repositories in our dataset. 
Hence, we tried to make the approach extendable 
for automatic topic recommendation in other software platforms as well.
Also for training the models, 
datasets were randomly split to avoid introducing bias.

\textbf{Construct validity} 
relates to theoretical concepts 
and use of appropriate evaluation metrics.
We have used standard theoretical concepts 
that are already evaluated and proved in academic society. 
Furthermore, we have carefully evaluated our results 
based on various evaluation metrics both 
for assessing multi-label classification methods and recommender systems.
Our results indicate that 
the employed approach has been successful 
in recommending topics for software entities.


\section{Conclusion} \label{sec:con}
Recommending topics for software repositories 
helps developers and software engineers 
access, document, browse, and navigate through repositories more efficiently.
By giving users the ability to tag repositories, 
GitHub made it possible for repository owners 
to define the main features of their repositories with few simple textual topics.
In this study, we proposed several multi-label classifiers 
to automatically recommend topics for repositories 
based on their textual information 
including their name, description, README files, wiki pages, and their file names.
We first employed rigorous text-processing steps 
on both topics and the input textual information.
We mapped $29K$ sub-topics
to their corresponding featured topics
provided by the GitHub.
Then we trained several multi-label classifiers 
including LR and DistilBERT-based models 
for predicting $228$ featured topics of GitHub repositories.
We evaluated our models both quantitatively and qualitatively. 
Our experimental results indicate that our models can suggest topics 
with high $R@5$ and $LRAP$ scores of $0.890$ and $0.805$, respectively. 
According to users' assessment, 
our approach can recommend on average $4.48$ correct topics 
out of $5$ topics and it outperforms the baseline.
In the future, we plan to take into account 
the correlation between the topics more properly. 
We also can exploit code analysis approaches to boost our approach. 

Furthermore, using the output of our approach, 
one can boost the techniques on possible applications of this work such as
finding missing topics
or categorizing repositories 
using the set of featured and mapped topics.

\begin{acknowledgements}
The authors would like to thank the participants 
who assessed the quality of the proposed approach.
We also like to thank 
Mahdi Keshani, Mahtab Nejati, and Alireza Aghamohammadi  
for their comments and help.
\end{acknowledgements}

\bibliographystyle{abbrv}
\bibliography{main}
\end{document}